%% file: journal.tex
\documentclass[acmsmall, screen]{acmart}

\usepackage[draft]{todonotes}   
\usepackage{booktabs} 
\usepackage{amsfonts}
\usepackage{multirow}
\usepackage{algorithm}
\usepackage[noend]{algpseudocode}
\usepackage{amsmath}
\usepackage{titlesec} 
\usepackage{paralist}

\usepackage{microtype}
\usepackage{sidecap}

\def\BibTeX{{\rm B\kern-.05em{\sc i\kern-.025em b}\kern-.08emT\kern-.1667em\lower.7ex\hbox{E}\kern-.125emX}}

\setcopyright{none}
\renewcommand\footnotetextcopyrightpermission[1]{}

\definecolor{yellow}{RGB}{247,198,3}
\definecolor{Blue}{RGB}{0,115,205}
\definecolor{Plum}{RGB}{148,0,211}
\definecolor{OliveGreen}{RGB}{85,107,47}
\definecolor{Red}{RGB}{220,20,60}

\setcopyright{acmcopyright}
\acmJournal{TOMM}
\acmYear{2020} \acmVolume{1} \acmNumber{1} \acmArticle{1} \acmMonth{1} \acmPrice{15.00}\acmDOI{10.1145/3396520}

\begin{document}

%
\title{Upgrading the Newsroom: \protect\\An Automated Image Selection System for News Articles}

%

\author{Fangyu Liu}
\authornote{Work done when FL was at LSIR, EPFL.}
\email{fl399@cam.ac.uk}
\affiliation{%
  \institution{Language Technology Lab (LTL), University of Cambridge}
  \city{Cambridge}
  \state{Cambridgeshire}
  \country{United Kingdom}
}

\author{R\'emi Lebret}
\email{remi.lebret@epfl.ch}
\affiliation{%
  \institution{Distributed Information System Lab (LSIR), EPFL}
  \city{Lausanne}
  \country{Switzerland}
}

\author{Didier Orel}
\email{didier.orel@tamedia.ch}
\affiliation{%
  \institution{Tamedia}
  \country{Switzerland}
}

\author{Philippe Sordet}
\email{philippes.ordet@tamedia.ch}
\affiliation{%
  \institution{Tamedia}
  \country{Switzerland}
}

\author{Karl Aberer}
\email{karl.aberer@epfl.ch}
\affiliation{%
  \institution{Distributed Information System Lab (LSIR), EPFL}
  \city{Lausanne}
  \country{Switzerland}
}


%

%

\input{contents/abstract.tex}
\maketitle
\input{contents/intro.tex}
\input{contents/related_work.tex}

\input{contents/dataset.tex}
\input{contents/method.tex}
\input{contents/recipe.tex}

\input{contents/experiment.tex}
\input{contents/limitation.tex}
\input{contents/conclusion.tex}
\input{contents/ack.tex}

\bibliographystyle{ACM-Reference-Format}
\bibliography{journal}

%









\newpage
\input{contents/appendix.tex}

\end{document}

%% file: contents/abstract.tex
\begin{abstract}
We propose an automated image selection system to assist photo editors in selecting suitable images for news articles. The system fuses multiple textual sources extracted from news articles and accepts multilingual inputs. It is equipped with char-level word embeddings to help both modeling morphologically rich languages, e.g. German, and transferring knowledge across nearby languages. The text encoder adopts a hierarchical self-attention mechanism to attend more to both key words within a piece of text and informative components of a news article. We extensively experiment our system on a large-scale text-image database containing multimodal multilingual news articles collected from Swiss local news media websites. The system is compared with multiple baselines with ablation studies and is shown to beat existing text-image retrieval methods in a weakly-supervised learning setting. Besides, we also offer insights on the advantage of using multiple textual sources and multilingual data.
\end{abstract}

%% file: contents/intro.tex
\section{Introduction}

\textbf{\textit{Motivation.}} Photo editors play a key role in the traditional workflow of the journalism industry. They are usually in charge of a huge image database that belongs to the journal and capable of picking suitable images for news in a very short time. With the development of information infrastructure and the internet, the collection and distribution of image data also have been much easier. Relying on human brains to memorize the database and find matches is becoming unsustainable. While the flood of data is hard for the human to process, it provides abundant examples that enable the training of data-hungry machine learning algorithms in a self-supervised manner. 


Here, we propose the task of News Image Selection. With tons of (online) news samples composed of the headline, lead, an image with a caption and the article body, we build a system that learns from these samples, takes in only texts then ranks image candidates for the news. The task resembles \emph{query based visual search} task like Google Image Search. Entering the deep learning era, novel models have been proposed for query based visual search. This line of works adopts a two-branch architecture where each branch is an encoder for image or text. Image and text are both encoded into vectors to be compared with some similarity metric in a joint space. By jointly modeling vision and language, the class of models are usually referred to as Visual Semantic Embeddings (VSE). However, things are different in several ways inside the Newsroom. First, instead of one line of text (in the form of phrases or keywords) as we input to Google Image Search, news articles contain multiple text sources with different features and characteristics. Unlike phrases or keywords, these text inputs are considerably longer, and most of them (image caption, lead and article body) are composed of complete sentences with clear structures. Also, they are written for readers, containing (subjective) discussions or arguments, not necessarily describing an image. Second, previous VSEs are usually trained on academic datasets like MS-COCO~\cite{lin2014microsoft} and Flickr30k~\cite{plummer2015flickr30k}, which are human annotated, containing clean and accurate labels. We will show that the state-of-the-art model on \cite{lin2014microsoft,plummer2015flickr30k} fails on News Image Selection. We also will discuss why and rationalize how training on a noisy real-world dataset has significantly changed how we should train and evaluate the models. 

\begin{figure}[t]
\vspace{-0.2cm}
\centering
\includegraphics[height=10cm]{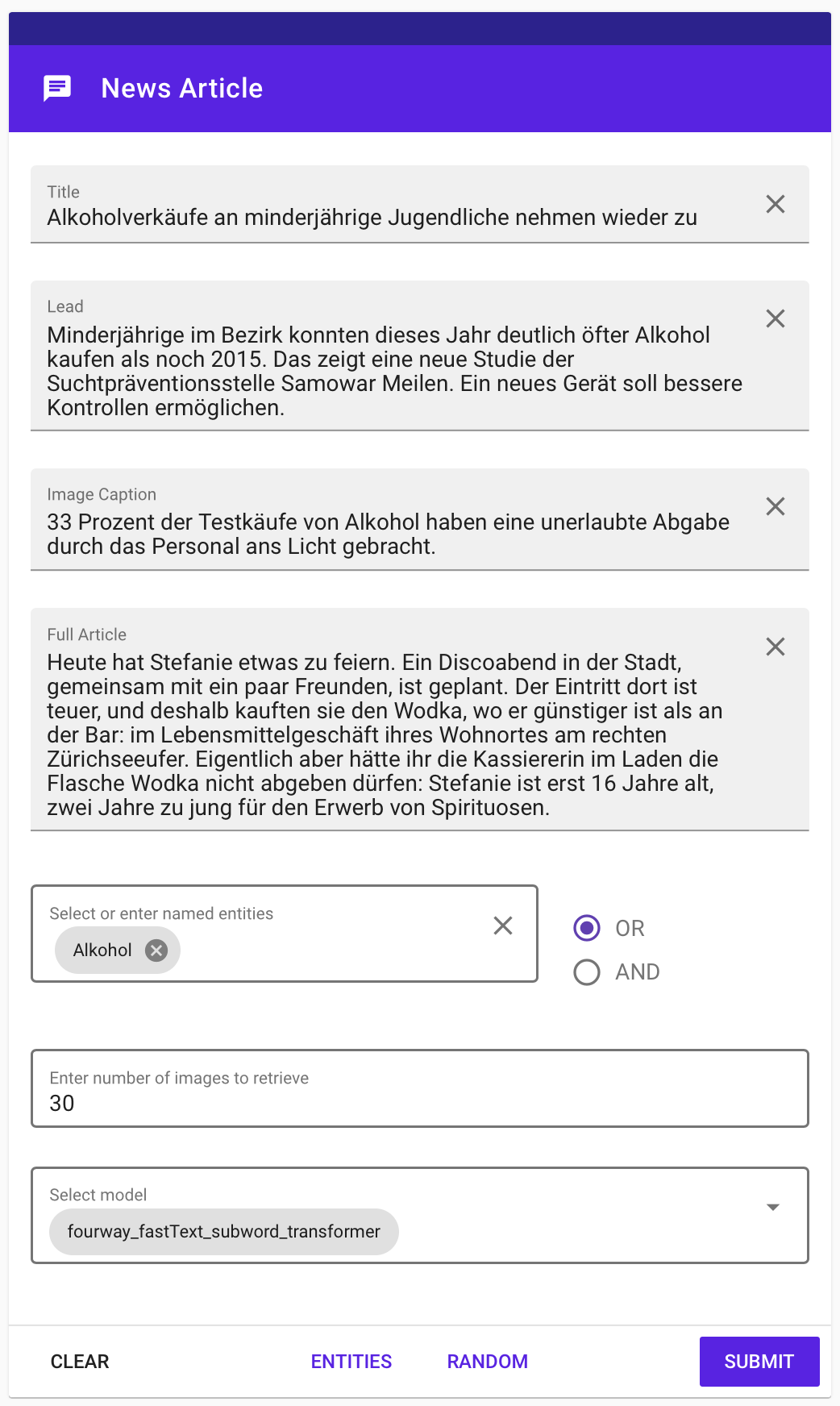}
\includegraphics[height=10cm]{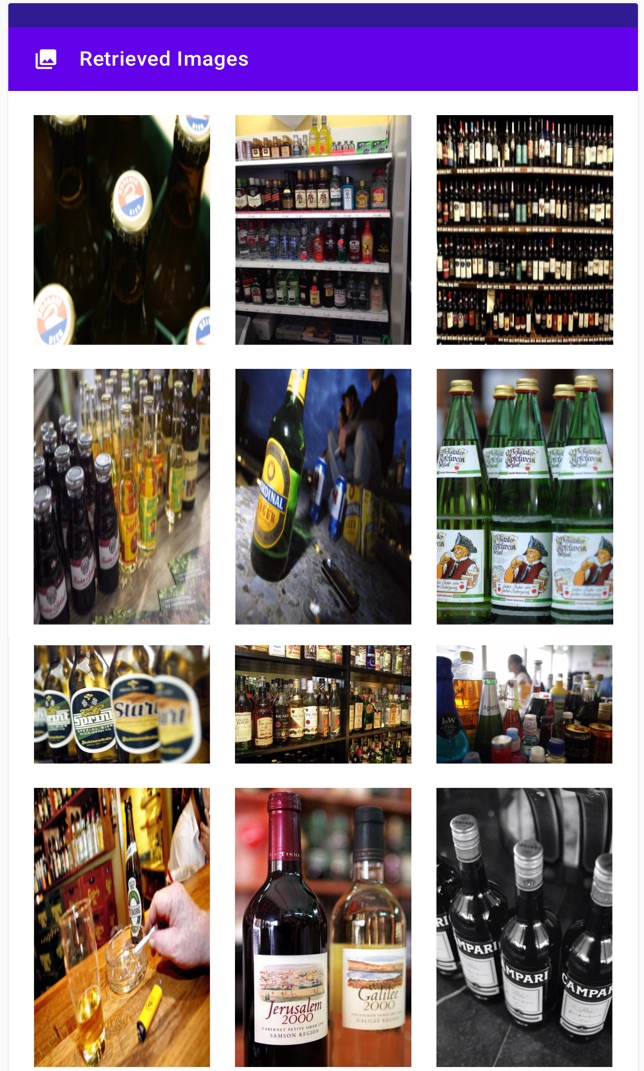}
\caption{A screenshot of the news image selection system's interface. Windows on the left-hand side hold the news article inputs with options to choose the desired entity(ies), the number of images expected and model choice while right-hand side lists the retrieved images in ranked order. This demo can be accessed at \texttt{https://modemos.epfl.ch/article}.\\Notice that the demo image dataset contains only limited published images. In the real world industrial setting, we extend this to a much larger full archival image database. }
\label{fig:screenshot-interface}
\vspace{-0.3cm}
\end{figure}



With the above analysis considered, we propose an image selection system for news that fuses multiple textual sources with a hierarchical attention mechanism. First, we conduct word-level attention within each textual source to select out the informative keywords. As regardless of the structure and layout of the article and sentences, the event or entity (in the form of keywords) the news talks about should be what to emphasize. Then, we apply component-level attention to decide how much to rely on each textual source. As in application, certain textual source would potentially be missing. Also, even if all modalities are complete, they contain different level of information that varies from news to news. Besides hierarchical attention, we also introduce the use of pre-trained subword embeddings and novel training techniques for multiple text sources. In the Experiment section, we will both qualitatively and quantitatively show how these improvements have made our image selection system to perform better than baselines.


We mainly focus on enhancing text encoder in this work as the \emph{de facto} image encoders - deep Convolutional Neural Networks (CNN) pre-trained on ImageNet~\cite{deng2009imagenet} yield high-quality image encodings and are robust image feature extractors in almost every vision-related task including Image Retrieval. Comparatively, text encoders in the text-image retrieval literature are less impressive and also not being paid enough attention to. Simple models like randomly initialized standard word embeddings plus single layer Recurrent Neural Network (RNN) or CNN were used since the start of this line of work~\cite{kiros2014unifying} (2014) and are still being used today~\cite{faghri2018vse++,liu2020hal} (2018 \& 2020). Our work would be one of the first ones to use n-gram subword embeddings and self-attention mechanism in the context of Image Retrieval.

\begin{figure}[h]
\vspace{-0.3cm}
\centering
\includegraphics[width=10.5cm]{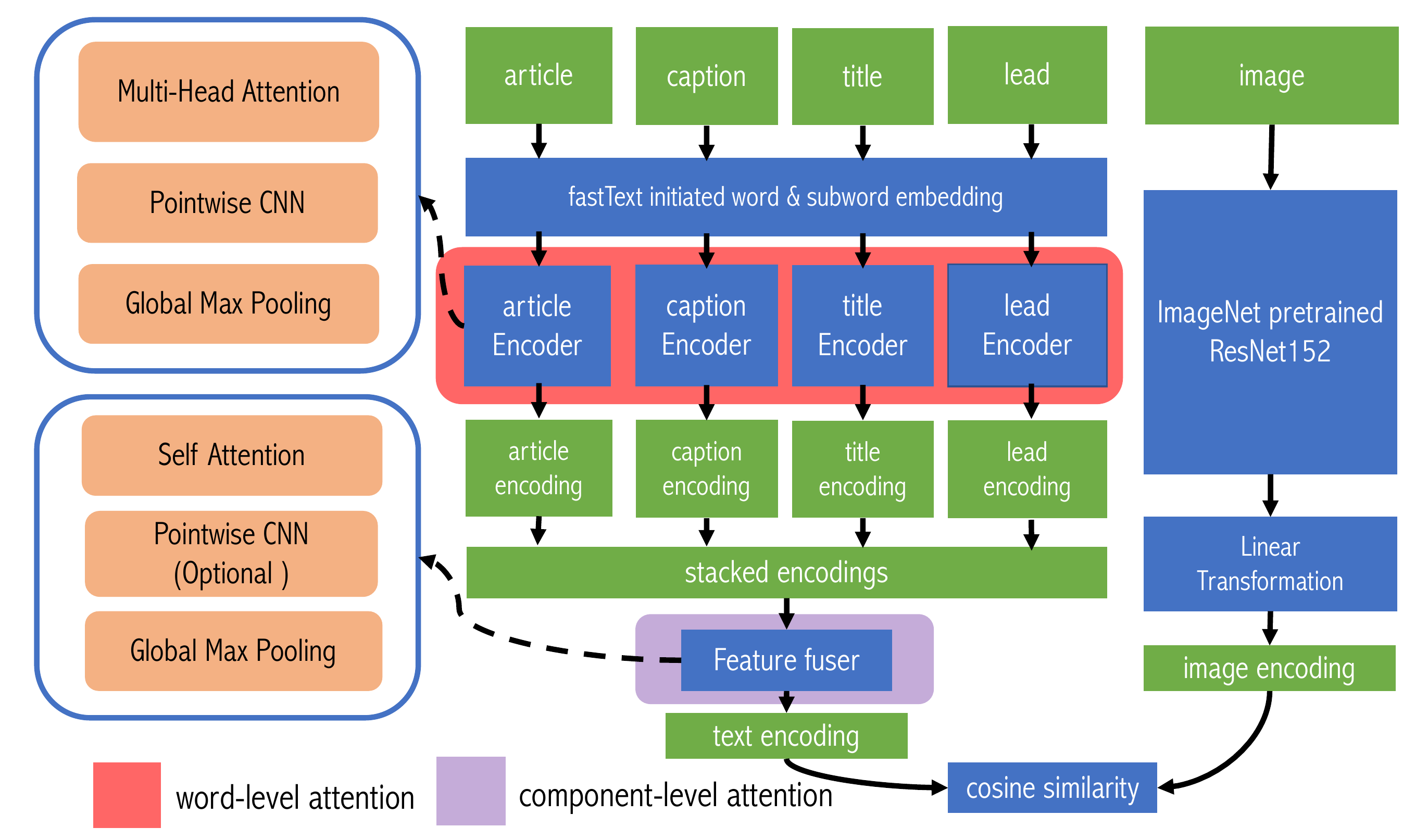}
\caption{Architecture of the multi-textual-source model. On the left branch, the four textual sources are input parallelly into word embedding. Embeddings of each source go into its corresponding text encoder. A feature fuser further combines encodings from the four textual sources into one single text encoding. On the right branch, a ResNet152 pre-trained on ImageNet encodes input image into a vector. The vector goes through a linear transformation and is then regarded as image encoding. Cosine similarity is used for computing rankings among text encodings and image encodings.}
\vspace{-0.3cm}
\end{figure}



In addition, there is another crucial aspect of Newsrooms - they (potentially) need to work multilingually, as we do here in Switzerland. It is also reflected in our database - it contains both German and French news articles. Simple searching methods based on keywords matching on metadata (which have been used in Newsrooms nowadays) clearly fail in the multilingual setting as only images attached to the specific language will be retrieved (not to mention that the quality of tagging in metadata is heterogeneous). We thus propose a language-agnostic model using aligned multilingual word embeddings. Being trained in both German and French, it outperforms models trained on monolingual data in both German and French text-image retrieval. Also, we explore the potential of transferring knowledge across language using our multilingual model. Starting with model weights trained on language A, we can achieve much better retrieval performance on language B by substituting only the word embedding and finetuning.

\textbf{\textit{Contributions.}} In this paper, we present a novel system for multimodal multilingual bidirectional retrieval with image and text as modalities. It is equipped with advanced language encoders inspired by latest advancements in machine translation \cite{bahdanau2015neural,vaswani2017attention,lample2017unsupervised} and char-level word embeddings \cite{bojanowski2017enriching}, considering subword information and utilizing attention mechanisms in different levels. The system benefits from fusing multiple textual sources and learns from multilingual training data, demonstrating outstanding performance in the News Image Selection task. We experiment with our system on a database collected from Swiss online news websites, larger than any existing public text-image dataset. We compare our model with various baselines on the database with ablation studies, showing that the proposed model is more suitable for News Image Selection than existing text-image retrieval methods. 

Our main contributions are as follows:

\begin{itemize}
\item Proposing the task of News Image Selection - a large-scale, multimodal, multilingual and weakly supervised machine learning task;
\item An attention-based image retrieval system fusing different sources of text inputs and accepting different languages;
\item A recipe for training multimodal multilingual models and transferring knowledge among languages, achieving better results in all languages on the task of text-image retrieval.
\end{itemize}

The rest of the paper is organized as follows: in Section \ref{sec:related}, we discuss related work; Section \ref{sec:dataset} describes our multimodal multilingual news database;  in Section \ref{sec:method}, we describe our multimodal bidirectional retrieval approach; in Section \ref{sec:recipe}, we introduce the ingredients for adapting the retrieval model to online news articles; in Section \ref{sec:exp}, we discuss our experiments and performance results; then we reflect the limitation of this work in Section \ref{sec:limitation} and conclude in Section \ref{sec:conclusion}.

%% file: contents/related_work.tex
\section{Related Works}
\label{sec:related}

\subsection{(News) Image Recommendation and Analysis}
The News Image Retrieval task has not been formally defined and well-studied before. There are, however, similar problems related to web (news) images being investigated in web content analysis community. 

Project DBpedia \cite{auer2007dbpedia} aims to extract structured information from Wikipedia. \citet{khalid2011framework} utilize this structured information to retrieve news images. \citet{Chatfield15} study visually searching large-scale video datasets for semantic entities specified by a text query.  \citet{feng2010topic,feng2010many,feng2013automatic} propose neural methods for automatically captioning news images. \citet{lin2018vizbywiki} build a system to retrieve relevant visualizations that already exist on the web, given a news article. However, \cite{auer2007dbpedia,khalid2011framework} are very out of date; \cite{Chatfield15} mainly focuses on short-query based searching which is more similar to Google-Image-Search-style visual search; \cite{feng2010topic,feng2010many,feng2013automatic}'s task is fundamentally different as it seeks a caption for an image but not the other way around; \cite{lin2018vizbywiki} is limited to retrieve only visualizations which already have context themselves.

\citet{ramisa2018breakingnews} propose a news text-image dataset called BreakingNews collected from English news media websites like \emph{BBC News}, \emph{The Guardian} and \emph{The Washington Post}. It contains around 100k news articles. Along with the dataset, \citet{ramisa2017multimodal} propose several tasks regarding multimodal news analysis like source detection, text illustration, and image caption generation etc.. However, they do not define any retrieval task, which would be what we are proposing. Also, the dataset we use is bilingual (German \& French) and significantly larger ($>$500k samples) than BreakingNews ($\approx$100k samples). We formally introduce our dataset in details in Section~\ref{sec:dataset}.

\subsection{Text-Image Retrieval and Visual Semantic Embeddings}

\emph{Siamese Network}~\cite{bromley1994signature} introduces the idea of using two identical networks to find a similarity or a relationship between two comparable things. It provides a basic framework for combining or comparing different modalities. Besides \cite{bromley1994signature} which is designed for discriminating signatures, later works use similar structures for face verification~\cite{chopra2005learning}, dimension reduction~\cite{hadsell2006dimensionality}, measuring semantic similarity~\cite{learning-deep-structured-semantic-models-for-web-search-using-clickthrough-data}, etc.
 
 Entering the era of deep learning, works emerged using this two-branch structure to connect both language and vision. \citet{frome2013devise} bring up the idea called \emph{Visual-Semantic Embedding (VSE)} to embed pairs of (text, image) data and compare them in a joint space. Later works extend \emph{VSE} for the task of text-image retrieval \cite{hodosh2013framing,kiros2014unifying,gong2014improving,ma2015multimodal,vendrov2015order,tsai2017learning,wang2018learning,wang2018joint,lee2018stacked,faghri2018vse++,liu2019strong,wu2019learning,li2019visual,liu2020hal}, which is similar to our work when using a single text source. Notice that text-image retrieval is different from generating novel captions for images, e.g. \cite{lebret2015phrase,Karpathy_2015_CVPR,feng2019unsupervised}, but to retrieve existing descriptive texts or images in a database. Another line of highly related work is called Story Visualization \cite{ravi2018show,chen2019neural}, which seeks to retrieve a sequence of images that describes a piece of text (composed of multiple sentences). Our News Image Selection task aims to recommend atomic images for news articles instead of retrieving sequential visualizations. 
 
 The text-image retrieval literature mainly validates models on small-scale well-labeled standard text-image datasets. We argue that the News Image Selection task differs from standard text-image retrieval not only in terms of its multi-sources of textual inputs (and potentially multilingualism) but more importantly, also it is put in the context of weakly supervised learning. \citet{faghri2018vse++} propose a hardest negative mining method used by almost all state-of-the-art systems across several academic datasets \cite{engilberge2018finding,lee2018stacked,wu2019unified,li2019visual}. However, it performs poorly on our data. In Section~\ref{sec:quant}, we will discuss in details that why it fails. And we would discuss in details how this weakly-supervised-learning setup has made a major difference when it comes to creating and evaluating models. We also conduct extensive experiments comparing our proposed model with previous works.

%% file: contents/dataset.tex
\section{Dataset}
\label{sec:dataset}


\subsection{Dataset Description}



Our dataset consists of 528,474 articles published between 2007-11-19 and 2018-05-08 on \emph{20 Minuten}, \emph{Der Bund}, \emph{Tages Anzeiger}, \emph{Berner Zeitung}, \emph{Z\"urichsee Zeitung}, \emph{Der Landbote}, \emph{Thuner Tagblatt}, etc. (German);
\emph{20 Minutes}, \emph{Le Matin}, \emph{24 Heures}, \emph
{Tribune de G\`eneve}, etc. (French) 
which are popular news websites in Switzerland. 
As seen in Figure~\ref{fig:components}, all news articles contain at least one image along with the four textual modalities, listed below by their level of abstraction from the highest to the lowest: 
\begin{compactitem}
    \item \emph{Headline}: text above a newspaper article, indicating its topic.
    \item \emph{Lead}: sentence capturing the attention of the reader and summing up the focus of the story.
    \item \emph{Caption}: basic information needed to understand the image and its relevance to the news\footnote{One may raise the concern that a caption is only identified after an image is selected and should not be used as part of a query. But actually, captions rarely directly describe their corresponding images and may as well be written before selecting an image.}.
    \item \emph{Body}: multiple paragraphs with quotes, details, and elaboration. 
\end{compactitem}

In the rest of the paper, we call a tuple of (image, caption, body, headline, lead) one sample where the first entry -- image -- means digital photos and the remaining four entries are the textual elements. There are four national languages in Switzerland: German, French, Italian, and Romansh. The two dominant languages being German and French\footnote{In 2015, the population of Switzerland was 63.0\% native speakers of German, 22.7\% French, 8.4\% Italian; and 0.6\% Romansh.}, our dataset contains 350,204 German samples and 178,270 French samples. 

\subsection{A Weakly-supervised Learning Task}

We view the News Image Selection task as a weakly-supervised learning problem for two reasons.

First, from the perspective of optimization, we have a relatively loose metric for \emph{good} retrievals. In standard text-image retrieval, for a specific text, the \emph{good} image in a database is usually unique and models aim to rank that image as high as possible. However, we do not rigorously require the exact positive image being ranked in the top. Instead, we expect a group of related (\emph{good}) images being ranked in the top area. As long as images in the top are meaningful, we actually do not care if the original corresponded image is among them or not.

Secondly, in terms of training data, we only have a weakly-labeled dataset. The data is weakly-labeled for reasons of three folds: 1) There is a discrepancy between the annotations and true semantic text-image links. For one text, more than one image could have been considered descriptive in the database. Contrarily, more than one text might be considered descriptive for one image. Though we are assuming the one-to-one correspondence between the two modalities to hold for training \& testing, the \emph{true} semantic links should be more heterogeneous in reality. 2) (text, image) pairs have very different levels of semantic correspondences. Some texts might be descriptive in a high level but would require extra knowledge to see their connections and some might even be very non-descriptive to each other as the selection of images is subjective to every photo editor. 3) In the multi-textual-source setting, one or more textual sources are sometimes missing, leaving the grounding signal to be weak and inconsistent.

Methodology-wise, the weakly supervised learning setting has affected our choice of learning objective and training strategy: 1) We opt for a loss function that resists inaccurate/false labels well, which is different from what has being used for most state-of-the-art models on MS-COCO. We justify our choice with quantitative evidence and extensively discuss the intuition in Section \ref{sec:quant}. 2) We introduce a \emph{random\_drop} regularization technique to accommodate missing textual source(s) during training, preventing the model from overly relying on specific textual input(s). We explain more of its effectiveness in Section \ref{sec:experiment-multi-text}.

%% file: contents/method.tex
\section{A Multimodal Bidirectional Retrieval System}
\label{sec:method}

In this section, we describe the details of our retrieval system. 
Given a single text source or multiple text sources as an input, we define our goal as retrieving the best matching images, or vice versa. We introduce our system that achieves this goal in three main modules:
\begin{compactenum}
    \item a bidirectional retrieval framework (Section \ref{sec:retrieval});
    \item an image encoder which represents the image input in a $d$-dimensional vector (Section \ref{sec:image_enc});
    \item a text encoder which represents the text input(s) in a $d$-dimensional vector (Section \ref{sec:text_enc}).
\end{compactenum}

\subsection{Bidirectional Retrieval Framework}
\label{sec:retrieval}

\begin{figure}
\vspace{-0.1cm}
\centering
\includegraphics[width=.8\textwidth]{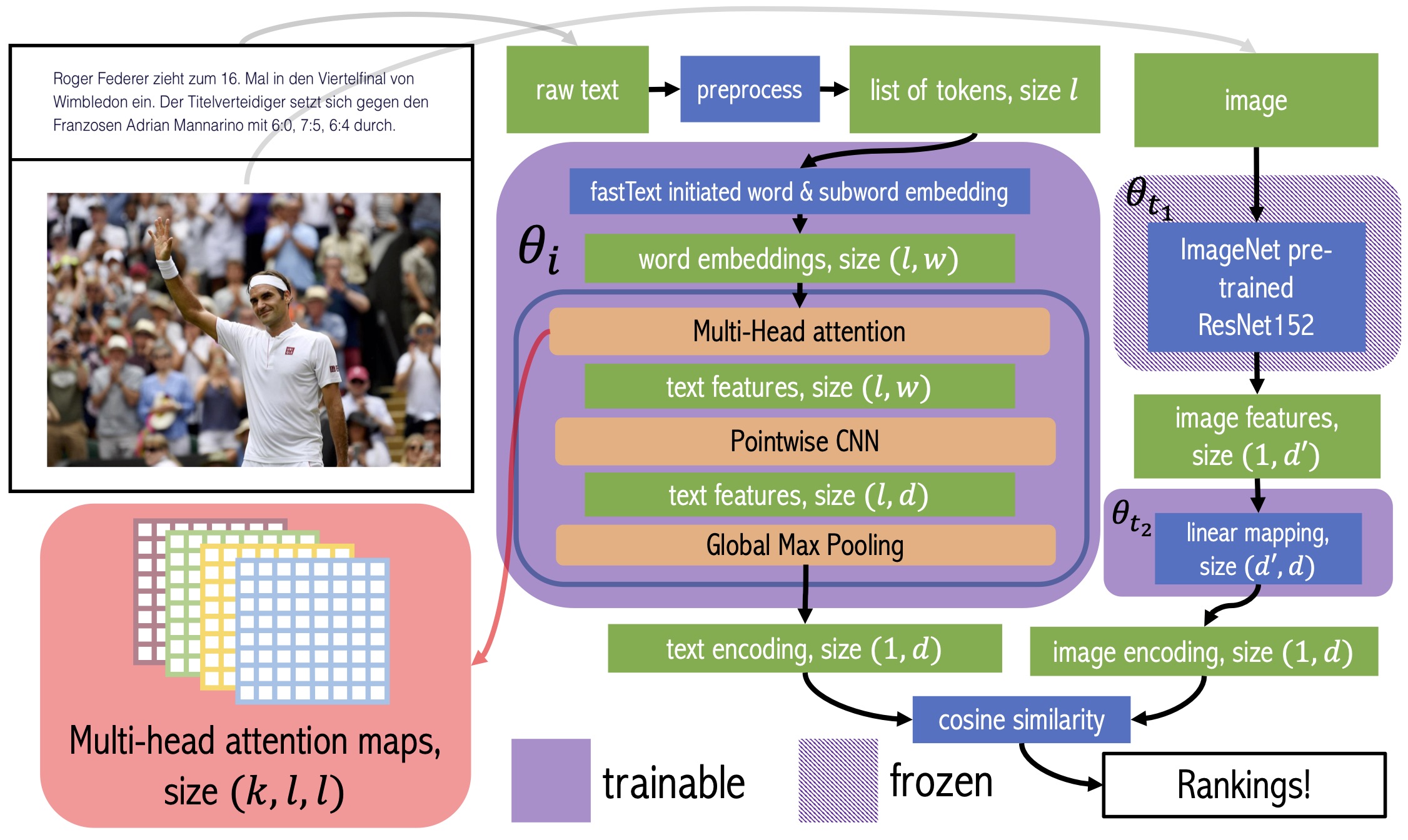}
\caption{This figure demonstrates a single-textual-source model's workflow in details. For a (text, image) pair, we input the text into the left branch and image to the right branch to get encodings for both modalities.
With one pair of text and image encodings, we can measure their similarity using cosine. Having obtained encodings for all the text and image samples, any combination of (text, image) can be measured and for any of text or image. We can then give a ranking of its closest counterpart in the opposite modality.}
\label{fig:unimodal}
\vspace{-0.5cm}
\end{figure}

\textbf{\textit{Model.}} Our bidirectional\footnote{It is called ``bidirectional'' as it can conduct both text-to-image and image-to-text retrieval.} retrieval framework consists of a text encoder and an image encoder. For a text-image pair $(t,i)$, the text encoder acts as a function $f$ which encodes $t$ into a vector representation $f(t; \theta_t)\in \mathbb{R}^{d}$; similarly, the image encoder encodes $i$ into another vector representation  $g(i;\theta_i) \in \mathbb{R}^{d}$, where $\theta_t,\theta_i$ are parameters of the two encoders and $d$ is dimension of the vector representations. To ensure that the encoded representations lie in a unit hyperspace, we further normalize them by $l^2$-norm. The text and image representations $\mathbf{x}_t$ and $\mathbf{x}_i$ would respectively be:
$$\mathbf{x}_t = \frac{f (t; \theta_t)}{\| f (t; \theta_t) \|_2}\in \mathbb{R}^{d}, \mathbf{x}_i = \frac{g (i; \theta_i)}{\| g (i; \theta_i) \|_2}\in \mathbb{R}^{d} .$$

The encoded and normalized text and image representations are of the same size and live in a joint space, convenient for us to compare their similarities. We define a simple inner product and obtain a scalar similarity score from it:
$$ s(i,t) = \langle \mathbf{x}_i,\mathbf{x}_t \rangle : \mathbb{R}^{d} \times \mathbb{R}^{d}\rightarrow \mathbb{R} .$$ 


\textbf{\textit{Training.}} Let $\theta$ denote all the model parameters that receive gradients. Let $I,T$ be all the images and texts samples respectively.
We optimize a margin based triplet ranking loss: 

\begin{equation}
\begin{split}
\mathcal{L_\textsc{Sum}} =
 \sum_{i\in I} \sum_{\bar{t}\in T\backslash \{t\}} [\alpha -s(i,t)+s(i,\bar{t})]_+ + \sum_{t\in T} \sum_{\bar{i}\in I\backslash \{i\}}  [ \alpha -s(t,i)+s(t,\bar{i}) ]_+,
\end{split}
\label{eq:loss}
\end{equation}
where $[\cdot]_+ = \max\{0,\cdot\}$; $\alpha$ is a preset margin; $t$ is the descriptive text for image $i$ and vice versa; $\bar{t}$ denotes non-descriptive texts for $i$ while $\bar{i}$ denotes non-descriptive images for $t$.

\subsection{Image encoder}
\label{sec:image_enc}
We use a Convolutional Neural Network (CNN) pre-trained on ImageNet~\cite{deng2009imagenet} as a feature extractor for images. Many off-the-shelf image classification neural nets like ResNet~\cite{he2016deep}, VGG~\cite{Simonyan14c} and DenseNet~\cite{huang2017densely} would be suitable for this need. For empirical reasons, we adopt ResNet152 as the image encoder architecture. We take the output of the layer before logits (which would be a vector of size $d'$) and send it into to a linear layer to be mapped to its required size $d$. So, the image encoder is composed of two parts: 1) a chopped pre-trained ResNet152 (frozen) and 2) a linear transformation (trainable). We denote the parameters of the chopped pre-trained ResNet152 and the linear transformation as $\theta_{i_1},\theta_{i_2}$ respectively where $\theta_{i_2}\in\mathbb{R}^{d'\times d}$ will simply be applied as a matrix multiplication. The image encoder as a whole is thus parameterized by $\theta_i = \{\theta_{i_1},\theta_{i_2}\}$. Since $\theta_{i_1}$ is fixed, trainable parameters in Equation~\ref{eq:loss} would result in $\theta = \{\theta_t, \theta_{i_2}\}$. The trainable and frozen parts of the model are also shown in Figure~\ref{fig:unimodal}. Denoting  $g' \in \mathbb{R}^{d'}$ the image representation extracted from a pre-trained CNN with ImageNet, the image encoder $g$ can be written as: $g(i;\theta_i) = g'(i;\theta_{i_1})\cdot \theta_{i_2}$.

\subsection{Text encoder}
\label{sec:text_enc}
Text encoder consists of a word embedding layer, a Multi-Head Attention layer, and a Point-Wise CNN. It resembles the Transformer architecture by \citet{vaswani2017attention} and utilizes the attention mechanism \cite{bahdanau2015neural} to softly align keywords in texts with latent visual objects.

We first walk through the pipeline of the text encoder as suggested in Figure~\ref{fig:unimodal}:

\begin{compactenum}
\item Assuming that the raw text $t$ has been pre-processed and tokenized, a word embedding layer maps every token in the sequence into a word embedding of size $w$. Given the sequence of $l$ tokens, we get a word embedding matrix of size $(l,w)$. 
\item The Multi-Head Attention takes the word embedding matrix in and produces a feature matrix of the same size.\footnote{Multi-Head Attention also outputs $h$ attention maps of size $(l,l)$ supposing that it is a $h$-head attention layer. We preserve the attention maps for the purpose of visualization. Details can be found in Figure~\ref{fig:attn-score-explained}.} 
\item The feature matrix is given as input to a Point-Wise CNN which produces an encoding of size $(l,d)$, where $d$ would be the preset text encoding feature size. 
\item A word-wise Global Max Pooling is then applied to obtain the final text representation $f (t; \theta_t) \in \mathbb{R}^d$.
\end{compactenum}

Appropriate word embeddings and an efficient attention mechanism are the two critical factors in the success of our task. When humans try to extract valuable information from news articles, they look for \emph{entities} only, rather than trying to understand complete sentences or structure of the article. This is where an attention mechanism plays its role: capturing \emph{entities}. However, in regular (pre-trained) word embeddings, \emph{entities} are more likely to be out-of-vocab than other words as they are usually context (language and location) specific. Without appropriate numerical representations of words, the attention mechanism would also fail. So, we enrich word embeddings with subword information to better encode entities even if they are out-of-vocab as a full word.
We now explain each part of the text encoder in detail.

\subsubsection{fastText n-gram word \& subword embedding}
We adopt the fastText Language Model pre-trained word and n-gram subword embeddings~\cite{bojanowski2017enriching}. fastText pre-trained word embeddings are considered as state-of-the-art word vectors. Also, n-gram subword embedding helps a lot for dealing with out-of-vocab words. 

 The general idea is that a word would be represented as the average of the vector representations of itself and its n-gram subwords. For example, considering the word \texttt{newsroom}. A regular word embedding layer looks it up in its vocabulary and return its vector representation if it's in-vocab or the vector for \texttt{<unk>} if it's out-of-vocab. But with our word \& subword embedding layer, it would be represented as the \textbf{average} of all word vectors of the set 

\begin{center}
\texttt{\{{$\underbrace{\color{Blue}\text{<newsroom>}}_{\text{original word}}$}, $\underbrace{\text{{\color{Red}<new},{\color{Red}news},{\color{Red}ewsr},{\color{Red}wsro},{\color{Red}sroo},{\color{Red}room},{\color{Red}oom>}}}_{3-\text{grams}}$,\\
$\underbrace{\text{{\color{OliveGreen}<news},{\color{OliveGreen}newsr},{\color{OliveGreen}ewsro},{\color{OliveGreen}wsroo},{\color{OliveGreen}sroom},{\color{OliveGreen}room>}}}_{4-\text{grams}}$, $\underbrace{\text{{\color{Plum} <newsr},{\color{Plum}newsro},{\color{Plum}ewsroo},{\color{Plum}wsroom},{\color{Plum}sroom>}}}_{5-\text{grams}}$\}}.\footnote{We use 3 to 5-grams in our model as it empirically works the best for German.}
\end{center}

The word has been given a more continuous representation in which morphology of words would be encoded. This is a simple but extremely powerful technique in our context as it would tolerate typos, as well as compound words, and could even help to transfer knowledge across nearby languages. We would discuss more these benefits of subword information in the Experiments section~\ref{subword-experiments}. 

\subsubsection{Multi-Head Attention}
\label{sec:att}

Multi-Head Attention has first been introduced to improve Neural Machine Translation (NMT) systems~\cite{vaswani2017attention}. While recent NMT systems are mainly based on RNN \cite{johnson2017google} or CNN~\cite{kalchbrenner2016neural,gehring2017convolutional} architectures, the original paper presents a pure attention-based architecture called \emph{Transformer} achieving state-of-the-art performance on English-to-German and English-to-French translation tasks. As our task does not require outputting a corresponding sequence (translation) with order and grammar, we removed two modules from the original architecture: 1) the positional encoding which memorizes the input sequence order; 2) the masks which prevent seeing future words.

Our Multi-Head Attention layer consists of $h$ parallel heads where every head is acting as a Self-Attention layer.

\paragraph{Self-Attention}

A Self-Attention layer helps the encoder look at other words in the input sentence as it encodes a specific word. It creates three vectors from each of the encoder's input vectors (in this case, the embedding of each word). So for each word, it creates a Query vector, a Key vector, and a Value vector. These vectors are computed by multiplying the word embedding matrix $E \in \mathbb{R}^{l \times w}$ by three matrices $W^Q,W^K\in \mathbb{R}^{w\times d_k}, W^V \in \mathbb{R}^{w\times d_v}$ that we trained during the training process. Explicitly, it uses the following equation:
\begin{equation}
     \text{SelfAttention}(E) =  \text{Softmax}\Big(\frac{(EW^Q)(EW^K)^T}{\sqrt{d_k}}\Big)(EW^V),
\label{eq:attn}
\end{equation}
where $d_k,d_v\in\mathbb{N}^+$ are hyper-parameters (we use $d_k = d_v= 64$ for our model). The output is of size $l\times d_v$.

\begin{SCfigure}
\centering
\vspace{-0.3cm}
\includegraphics[width=8cm]{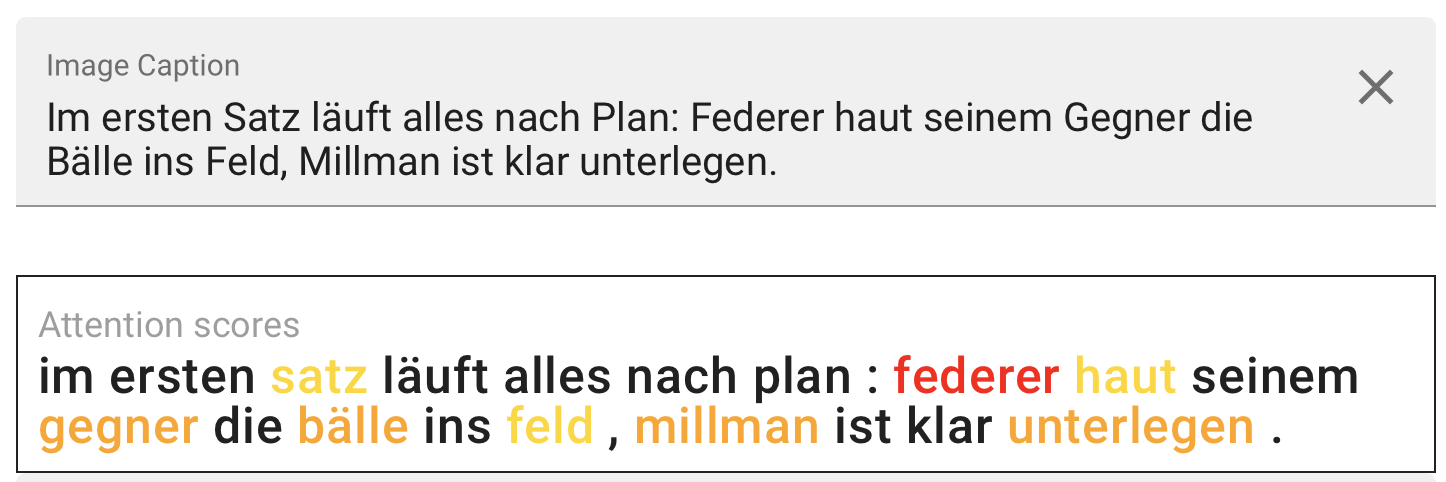}
\caption{A screenshot of visualizing attention scores in our web application. (Best seen in colors, attention decreases in the order of \textbf{\color{Red}red}, \textbf{\color{orange}orange}, \textbf{\color{yellow}yellow} and \textbf{black}.)}
\label{fig:screenshot-attn-score}
\vspace{-0.3cm}
\end{SCfigure}

For simplicity, we further denote $Q = EW^Q\in \mathbb{R}^{l\times d_k}, K = EW^K \in \mathbb{R}^{l\times d_k}, V = EW^V \in \mathbb{R}^{l\times d_v}$ where $Q,K,V$ are called \emph{query}, \emph{key} and \emph{value} respectively. Notice that with \emph{query}-\emph{key} matrix $QK^T\in \mathbb{R}^{l\times l}$ we could already do a weighted sum on the input. We further scale it with $\sqrt{d_k}$ and send it to $\text{Softmax}$ function for obtaining weight maps with row sum and column sum that equals to $1$:
\begin{equation}
M = \text{Softmax}\Big(\frac{QK^T}{\sqrt{d_k}}\Big).
\label{eq:attn-3}
\end{equation}
We call $M\in \mathbb{R}^{l\times l}$ an attention map as it decides dependencies of the output vectors on the input vectors. The map is preserved to be used for visualization like the toy example in Figure~\ref{fig:screenshot-attn-score}. We discuss more how attention maps can be used to interpret model's behavior in Experiments section~\ref{query-system}.

\paragraph{Multi-Head}

For a $h$-Head Attention (we used $h=6$), it concatenates outputs of $h$ independent Self-Attention layers and applies a linear transformation as suggested below:

\begin{equation*}
\begin{split}
     \text{MultiHead}(E) = [\text{head}_1,\cdots,\text{head}_h]W^O + E\\
     \text{where } \text{head}_i = \text{SelfAttention}_i(E)
\end{split}
\label{eq:multihead}
\end{equation*}
where $W^O\in\mathbb{R}^{hd_v \times d_{\text{model}}}$ are the weights of a linear transformation to be trained (we use $d_{\text{model}} = w$ for our model) ; $[\cdot]$ denotes the concatenation. It is worth mentioning that every $\text{head}_i$ is computed from its exclusive $\text{SelfAttention}_i$ function, which means that there are $h$ $\text{SelfAttention}$ modules working in parallel with their own independent weights. The Multi-Head Attention layer also adds its input directly to the final output which forms a residual connection.

\subsubsection{Point-Wise CNN}
The output of the Multi-Head Attention is sent into a Point-Wise CNN where kernels are all of size $1$:
\begin{equation*}
\begin{split}
    \text{PointWiseCNN} (A)=  \text{ReLU}(AW_1 + b_1)W_2 + b_2,
\end{split}
\end{equation*}
where $A\in \mathbb{R}^{l\times d_{\text{model}}}$ is the output of the Multi-Head Attention layer; $W_1\in \mathbb{R}^{d_{\text{model}}\times d_{\text{hidden}}}, W_2 \in \mathbb{R}^{d_{\text{hidden}}\times d}$ are weights to be trained ($d_{\text{hidden}}$ is a hyper-parameter, $d$ is the vector dimension of the final text encoding, we use $d_{\text{hidden}}=2048,d=1024)$; $b_1,b_2$ are biases. We use a rectified linear unit (ReLU)~\cite{glorot2011deep}  as activation function.

\subsubsection{Global Max Pooling}
After the Point-Wise CNN layer, we obtain a matrix of size ($l\times d$). We then apply a Global Max Pooling on the last dimension of this matrix to get the final representation of the text input $t$, $\mathbf{x}_t \in \mathbb{R}^{d}$. 

%% file: contents/recipe.tex
\section{Special Ingredients for News Articles}
\label{sec:recipe}


Different from standard Image Retrieval, News Image Selection task has some special ingredients: it incorporates multiple text modalities (image caption, article body, headline, lead) and sometimes required to accept multiple input languages. 

In the following two subsections, we explore the use of special ingredients in our news articles recipe. We aim to build models that, first, utilize more context (textual modalities) to do better searching; second, learns from multiple languages and accepts multilingual inputs.

\subsection{The More the Better - Textual Sources Fusion}

\begin{SCfigure}
\vspace{-0.3cm}
\centering
\includegraphics[width=9.0cm]{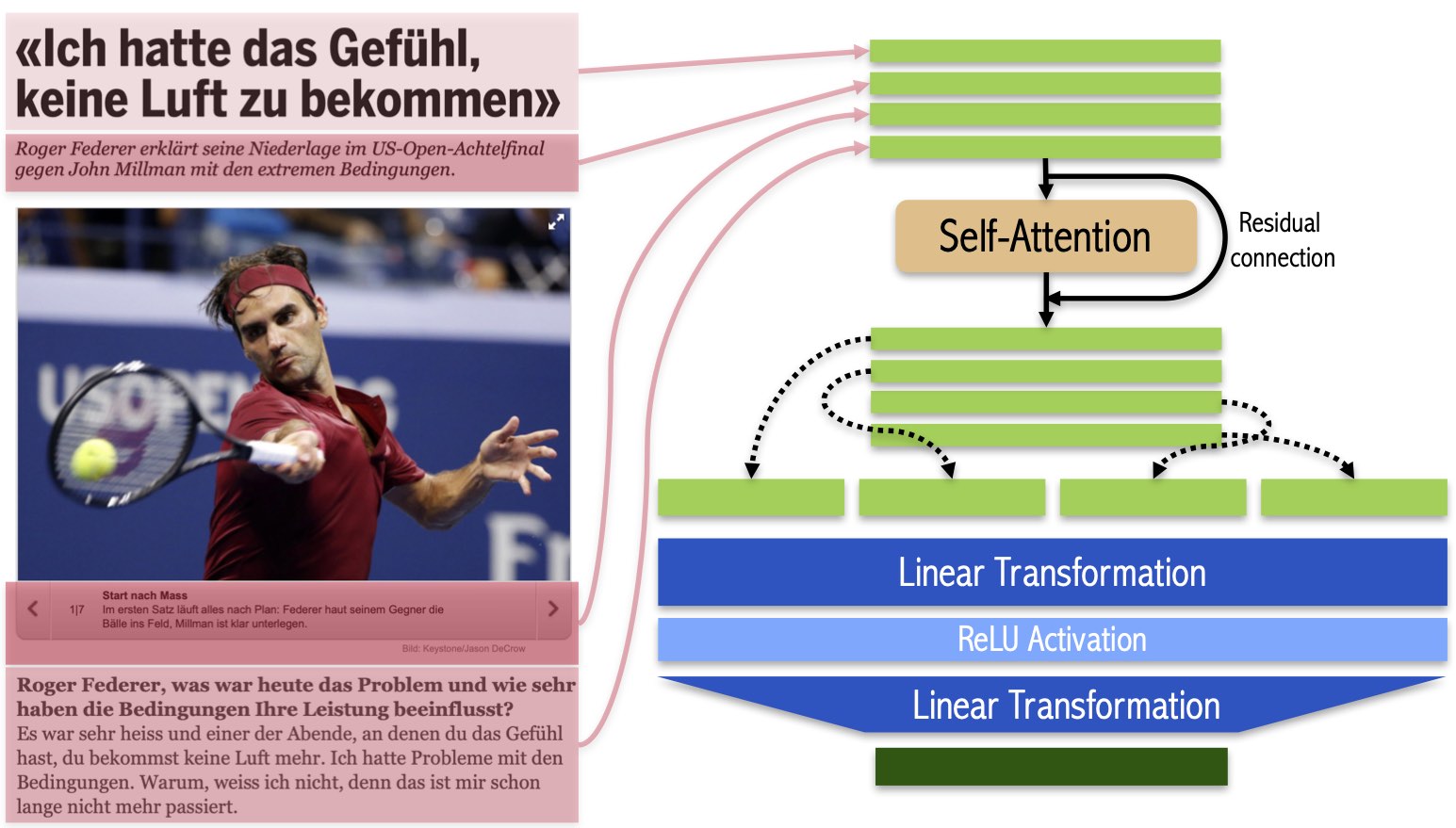}
\caption{For each text source input, the output of the text encoder is stacked together and provided as input to the Self-Attention layer, which decides how much attention is being paid to each textual source. Then the output is split into vectors and concatenated to be passed into two linear layers with a ReLU activation in between. }
\label{fig:components}
\vspace{-0.3cm}
\end{SCfigure}

The dataset described in Section \ref{sec:dataset} provides more than one text source, we make use of them. To fuse these text sources, we use more attention. Inside each text source, we've already had Multi-Head Attentions to conduct word-level attention, as described in Section \ref{sec:att}. But viewing from a higher level, different components of a news would contain different amount of information varying from news to news. Sometimes it's the article that tells more about the content of the image and sometimes it might be the lead, etc. We thus propose a \emph{Feature Fuser} that combines fully-connected layers with another component-level attention deciding the attention to put on each textual source as suggested in Figure~\ref{fig:components}.

Our \emph{Feature Fuser} achieves this goal through the following steps:
\begin{compactenum}
    \item obtains a $d$-dimensional feature vector for each of the $4$ different text sources;
    \item stacks these features together into one feature matrix of size $4\times d$;
    \item applies a Self-Attention (or Single-Head Attention) on it using the exact same equation for word-level attention in Equation~\ref{eq:attn};
    \item reshapes the resulting feature map of size $4\times d$ into $4d$-dimensional global feature vector by concatenating across its first dimension;
    \item feeds the global feature vector into two linear layers with a ReLU activation in between\footnote{Notice that the first linear transformation is of size $(4d,4d)$ while the second is $(4d,d)$.}.
\end{compactenum}
The final $d$-dimensional vector is regarded as the text representation of all four text sources and is used to compute similarities with image encoding.

\subsection{Rebuilding a (very small) Tower of Babel - Multilingual Model}

In this section, we attempt to overstride language barriers in text-image retrieval, demonstrating the potential of building language agnostic models and transferring knowledge across languages. We first briefly discuss how our multilingual setup is different from previous ones in Section~\ref{sec:no_p_c}. Then we introduce two ways of conducting transfer learning in Section~\ref{sec:14all} and \ref{sec:all41}.

\subsubsection{Transfer Learning without Parallel Corpus}
\label{sec:no_p_c}
In human brains, semantic concepts regardless of language are likely to be grounded similarly in the perceptual system~\cite{pecher2005grounding}. Recent works in the natural language understanding community have pointed out that by grounding multilingual data visually, a model can yield improved bilingual text representations and achieve better results in visual-language tasks including text-image retrieval ~\cite{rajendran2016bridge,gella2017image,rotman2018bridging}.

However, these proposed methods require parallel text corpus from different languages along with visual anchors, i.e., a dataset composed of (image, text of lang. A, text of lang. B) samples, where text of lang. A and lang. B describes a common image. Although there are indeed human-labeled academic datasets like this as mentioned in Section~\ref{sec:dataset}, they rarely exist in real world settings. 

In our setup, without parallel corpus, to still benefit from multilingual data, we choose to start with already aligned word embeddings. Instead of relying on multilingual data to align word embeddings, the model can focus on discovering connections of language and visual representations.

\subsubsection{One for All}
\label{sec:14all}
Wouldn't it be nice to have one model that understands all languages? In our multilingual context, it is very ideal to have a language agnostic model as people may search with different languages on a common image database. We achieve this by utilizing fastText \emph{MUSE} word embeddings~\cite{lample2017unsupervised,lample2018word,joulin2018loss}. In \emph{MUSE}, word embeddings of different languages are aligned to have alike shapes and structures in hyperspace, which means for tokens of similar meanings in different languages, \emph{MUSE} outputs similar numerical representations. This fact bridges data from different languages. 

In our image retrieval context, words with identical meanings in different languages would be grounded by similar images as vision is universal. In terms of the text encoder, it means: 1) from above, it receives uniform numerical signals of word vectors that do not depend on language; 2) down below, these language agnostic signals would always be grounded by images with universal  semantic meanings.

\begin{SCfigure}
\vspace{-0.3cm}
\centering
\includegraphics[width=7cm]{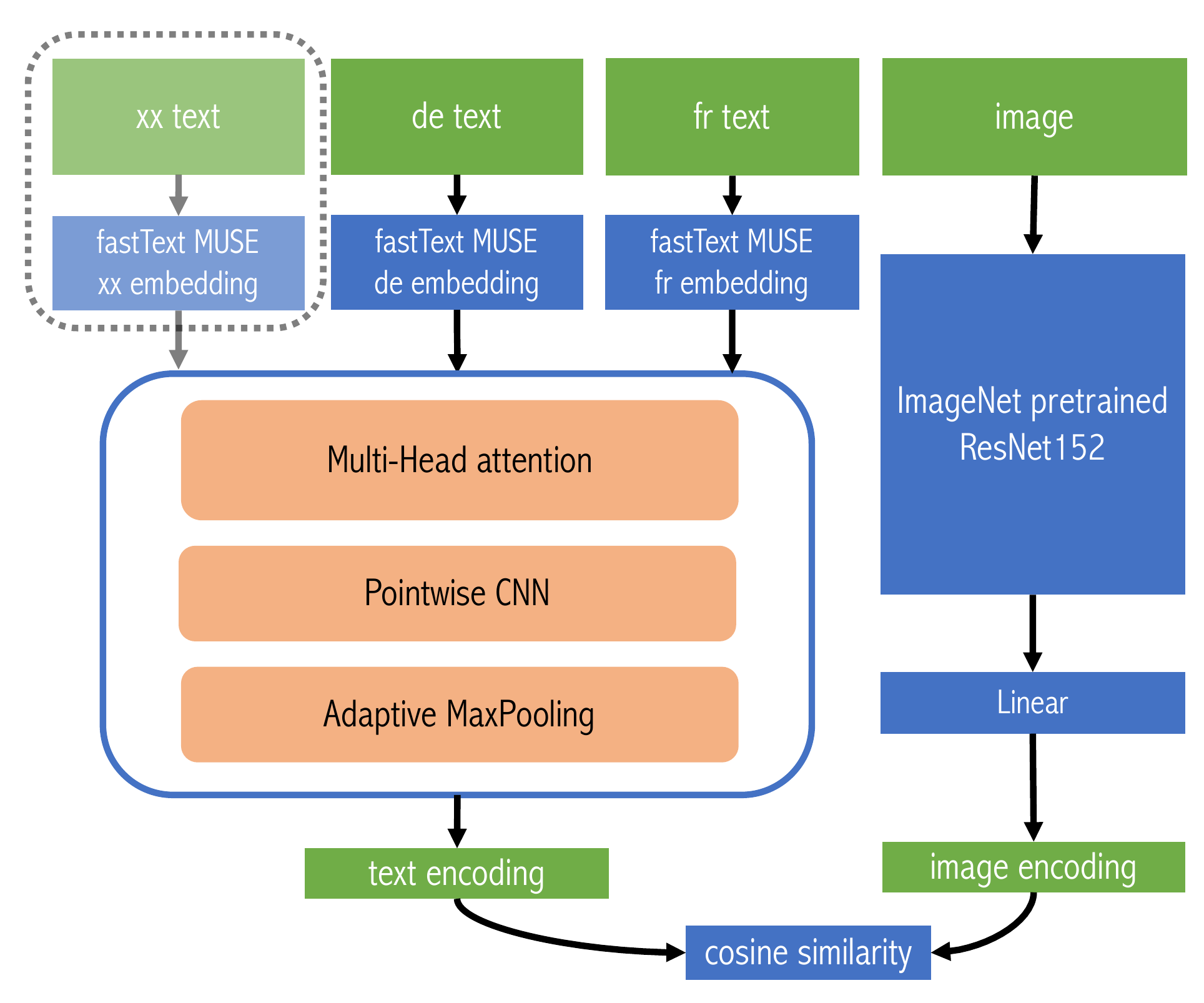}
\caption{\emph{One} model \emph{for All} languages. Though the model is only trained on German and French data, it is capable of taking other languages sources with its corresponding \emph{MUSE} word embeddings.}
\label{fig:one-for-all}
\vspace{-0.3cm}
\end{SCfigure}

Based on the above we propose the model in Figure~\ref{fig:one-for-all} which can take text inputs from different languages at the same time within one epoch in training. When the input language is changed, the word embedding switches to the corresponding language. To prevent the aligned hyperspaces from being broken, we freeze the word embeddings of all training languages.

\subsubsection{All for One}
\label{sec:all41}
Deep neural nets are data-hungry models. Imagine the scenario when we want to build a model for language B, but would like to also benefit from data in language A.

\begin{figure}[h]
\vspace{-0.2cm}
\centering
\includegraphics[width=12cm]{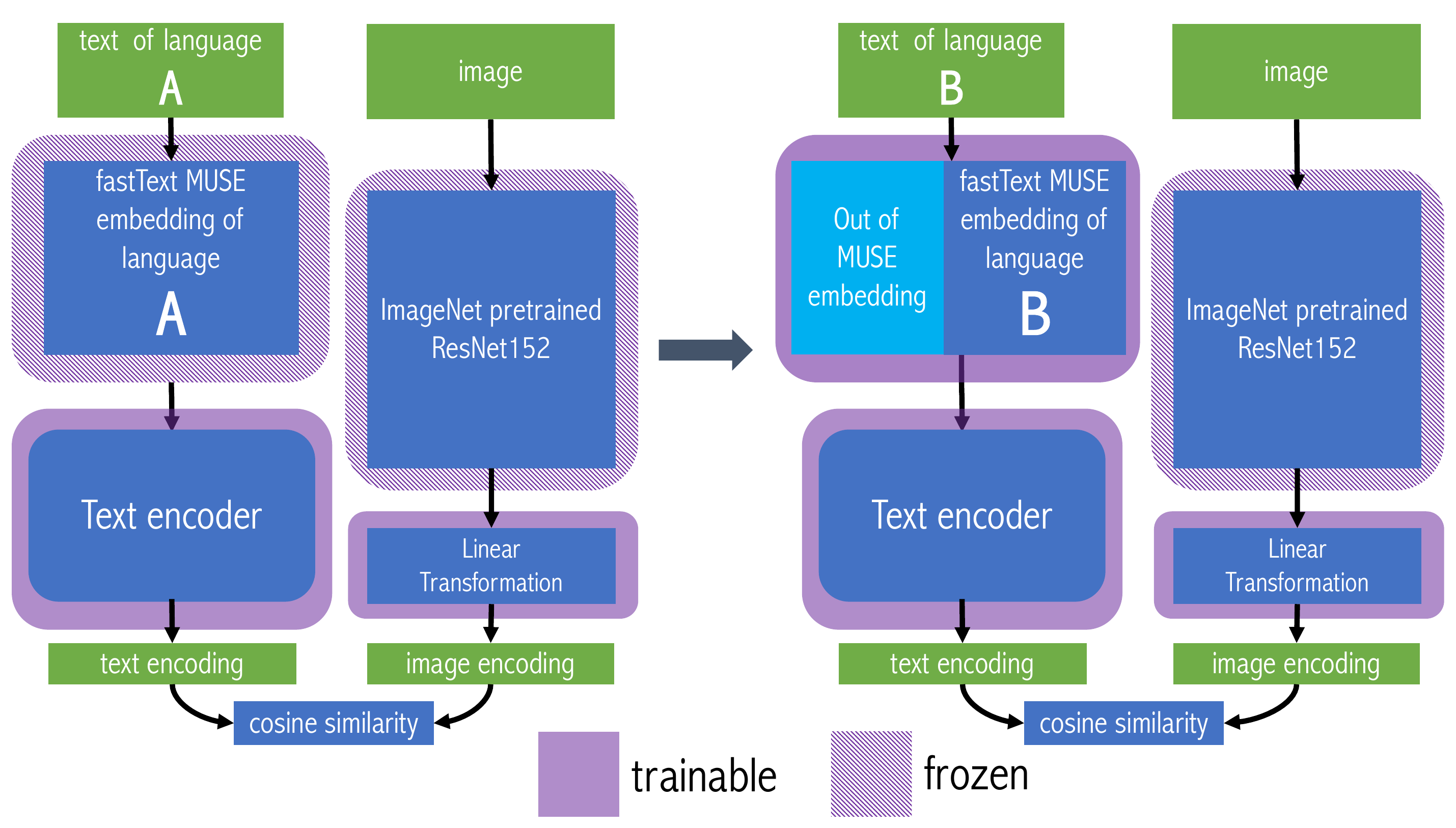}
\caption{Utilize data in \emph{All} languages \emph{for} training \emph{One} model.}
\label{fig:all-for-one}
\vspace{-0.2cm}
\end{figure}

We propose a recipe for transferring knowledge across languages: 1) Initialize word embeddings with fastText \emph{MUSE} weights for language A. Freeze word embeddings. Train model on language A; 2) Substitute the \emph{MUSE} word embeddings with language B's. Add vocabularies that are missing in \emph{MUSE} word embeddings. Preserve weights in other parts of the model. Train the model on language B, fine-tuning the word embeddings. This process can be visualized in Figure~\ref{fig:all-for-one}.

Low-resource languages can benefit the most from knowledge transfer. With this recipe, we can pre-train our model on large datasets of high-resource languages then fine-tune it on our own language of interest.

%% file: contents/experiment.tex
\section{Experiments}
\label{sec:exp}

\begin{table*}[h]
\small
\renewcommand{\arraystretch}{0.9}
\setlength{\tabcolsep}{3pt}
\caption{Comparison between our model with ablation studies (line 1.6-1.11) and baseline models (line 1.1-1.5) in the single-textual-source setting. Recall@10 is reported here. ``-'' means unable to converge.}
\label{table:unimodal}
\centering
\begin{tabular}{clcccccccccc}
\toprule
\multirow{2}{*}{\#} & \multirow{2}{*}{ {\bf Models} } & \phantom{abc} & \multicolumn{4}{c}{\bf image to text} & \phantom{abc} & \multicolumn{4}{c}{\bf text to image} \\ \cmidrule(l){4-7} \cmidrule(l){9-12}
& & & caption & body & headline & lead & & caption & body & headline & lead \\   \midrule
1.1  & \textsc{Vse++}~\cite{faghri2018vse++}   & & 39.7 & 25.2 & 17.8 & 27.3 &  & 41.2 & 26.7   & 17.1 & 26.7 \\
1.2 & \textsc{Vse} \cite{kiros2014unifying}   & & 42.4 & 32.6 & 22.2 & 29.2 & & 42.0 & 32.8 & 21.8 & 28.6 \\
1.3 & \textsc{Order} \cite{vendrov2015order}   & & 38.9 &  22.1 & - & - & & 39.3 & 22.6 & - & -\\
1.4 & \textsc{Chain} \cite{wehrmann2018cvpr}   & & 22.2 & 10.3 & 16.6 & 14.5 & & 20.7 & 11.7 & 16.9 & 14.4 \\
1.5 & \textsc{Hal} \cite{liu2020hal} & & 49.8 & 32.2 & 23.3 & 33.1 & & 49.9 & 32.3 & 22.6 & \textbf{33.5} \\
\midrule
1.6 & \textsc{Vse} + \emph{wiki} & & 48.3 &   35.8 &   23.0     &  31.7 & & 48.1 & 36.4 &   22.7 & 32.5 \\
1.7 & \textsc{Vse} + \emph{cc-wiki} & &  45.9 &  37.1   &  22.1 & 33.6 & &  46.2   &  \textbf{37.7}   &   22.3  &   32.4 \\
1.8 & \textsc{Vse} + \emph{wiki-subword} & & 52.8   & 34.5   & \textbf{24.9} & 32.1 & & 51.4 & 34.0 & \textbf{23.9} & 31.2 \\
1.9 & Ours (w/o attention) & & 46.2 &   33.8 &   22.7  &  32.2      & &   46.5 & 34.7 &  22.7  &  32.4 \\
1.10 & Ours  (w/o \emph{wiki-subword}) & & 45.8 & 28.2 &  20.2 & 26.5 & & 45.2 & 28.3 &  18.8 &  24.8 \\
1.11 & Ours (full model) & & \textbf{54.2} & \textbf{37.2} & 24.2 & \textbf{35.4} & & \textbf{52.0} & 36.8 & 22.9 & 32.2 \\ \bottomrule
\end{tabular}
\end{table*}
In this section we present experimental results in the order of single textual source; multiple textual sources; multilingual model. Before diving into them, we first introduce some general of model configuration, training \& evaluation schemes.

\textbf{\textit{Implementation, Training and data configurations.}} 
We implemented our system using PyTorch and referenced open source projects like~\cite{faghri2018vse++,gehring2017convolutional,rush2018annotated}. The models are trained on two NVIDIA TITAN X Graphics Cards. All single-textual-source models are trained for $30$ epochs starting with a learning rate of $0.001$ decaying $5$ times for every $10$ epochs. Training details of multi-textual-source models could be found in Section~\ref{sec:experiment-multi-text}. All models are optimized using the Adam optimizer~\cite{kingma2014adam}. The full dataset is split into train set and validation set. For data of each language, we randomly picked 1.2k samples as the validation set for that language. All rest are considered as training samples. Quantitative results in all tables come from evaluations conducted on these validation sets.

\textbf{\textit{Word Embedding Configurations.}} 
We used four kinds of pre-trained word vectors in experiments:
\begin{compactitem}
    \item fastText Wikipedia word embedding (\emph{wiki})~\cite{bojanowski2017enriching};
    \item fastText Wikipedia word \& subword embedding (\emph{wiki-subword})~\cite{bojanowski2017enriching};
    \item fastText Common Crawl \& Wikipedia word embedding (\emph{cc-wiki})~\cite{grave2018learning};
    \item fastText Multilingual Unsupervised and Supervised Embeddings (\emph{MUSE})~\cite{lample2017unsupervised,lample2018word,joulin2018loss}.
\end{compactitem} 
\emph{wiki-subword}, \emph{wiki}, and \emph{MUSE} are all pre-trained on (German or French, depending on the language of training model) Wikipedia while \emph{cc-wiki} is pre-trained on both Common Crawl and Wikipedia. \emph{wiki-subword} and \emph{wiki} are essentially the same word embedding except that \emph{wiki-subword} also considers subword information. Though \emph{cc-wiki} is pre-trained on more data, it contains a smaller vocabulary than \emph{wiki} and \emph{wiki-subword}. Results in Table~\ref{table:unimodal} (\emph{line 1.3, 1.4}) suggests that their performances are comparable while \emph{wiki} does better on image caption, which is the textual source we care about the most as it has the richest information for retrieving images. We thus conduct all other experiments on \emph{wiki} and \emph{wiki-subword}. \emph{MUSE} is only used in the multilingual setting. We further introduce \emph{MUSE} in Section~\ref{sec:exp-multilingual}.

\textbf{\textit{Metrics.}} 
We evaluate models by the commonly used Recall at $K$ (i.e. R@$K$) metric in Information Retrieval:
$$\text{R}@K = \frac{\text{\# of queries that the groundtruth item is ranked in top $K$}}{\text{total \# of queries}}.$$
And we use this metric for both text-to-image and image-to-text retrieval.
In single-textual-source setting, we report only Recall at top 10 retrievals (R@10). In our News Image Selection task (and also weakly supervised setting), we do not expect the model ranks the one single positive image as the high as possible, but to rank all related images in the top range. For such reason, we consider R@10 as our most relevant metric rather than R@1 or R@5. When fusing multiple textual sources, the best model achieves nearly $70\%$ for R@10, we thus include R@1, R@5 and also Median rank (Med r, the lower the better) to evaluate model from multiple angles.

\subsection{Models with One Single Textual Source}

In this section, we experiment single-textual-source models on German data, including all textual sources: caption, body, headline, and lead. We demonstrate experiment results in three perspectives: \emph{Quantitative results}, \emph{Multi-Head Attention as a query system} and \emph{Subword information matters}.

\subsubsection{Quantitative results}
\label{sec:quant}
We compare our model with five baseline models: 1) \textsc{Vse} \cite{kiros2014unifying}; \textsc{Vse++}~\cite{faghri2018vse++}; 3) \textsc{Order} \cite{vendrov2015order}; 4) \textsc{Chain} \cite{wehrmann2018cvpr}; 5) \textsc{Hal} \cite{liu2020hal}. 
According to prior works \cite{faghri2018vse++,wehrmann2018cvpr,liu2020hal} and our own experiments, an ImageNet pre-trained ResNet152 yields the best image encodings in our context. To be consistent, we adopt ResNet152 as image encoder to compare all text encoder architectures. We conduct grid search to tune the hyperparameters used in loss functions for each baseline. The presented results are produced using hyperparameters with the best performance.

Overall, our model (\emph{line 1.11}) demonstrates improvements comparing to all baseline models under all textual sources and in both text-to-image and image-to-text task.  Among baseline models, the recently proposed \textsc{Hal} performs the best across almost all textual sources. It uses a novel loss function and produce state-of-the-art results on MS-COCO (without using object bounding box labels). We analyze \textsc{Hal} in more details in the coming section of comparing loss functions. Following \textsc{Hal}, \textsc{Vse} remains a very strong baseline despite having succinct model architecture, achieving higher scores than \textsc{Vse++} and \textsc{Order}. In our industrial setting, models that are sensitive to noise are very hard to train and can result in poor performance or even failure of convergence (e.g. when training \textsc{Order} on headline \& lead data, line 1.3). We analyze more on this in the later section of comparing loss functions. \textsc{Chain} fails significantly, producing the worst results under all textual sources. We suspect it is due to \textsc{Chain} uses a char-level text encoder and does not explicitly have a word embedding module. The benefit from using word embeddings over char embeddings is particularly prominent when the vocabulary size is large such as in our dataset (>100k) as opposed to MS-COCO (10K).

\textbf{\textit{Ablation studies.}}  We claim that our model's outstanding performance mainly benefit from two modules: 1) subword embeddings and 2) multi-head attention. \emph{line 1.9} in Table~\ref{table:unimodal} represents our model without the multi-head attention module, in which case the text encoding is produced with subword embeddings followed by a point-wise CNN. The model's performance drops significantly comparing to the full model (\emph{line 1.11}) and is also worse than \textsc{Vse} equipped with subword embeddings (\emph{line 1.8}). \emph{line 1.10} lists results of our model without the subword embeddings. The model produces even lower scores. Only when both modules are combined, our model achieves significantly better results than baselines.

\textbf{\textit{Why Attention?}}
Here we try to interpret why the transformer-based attentional encoder performs better than RNNs like LSTM~\cite{hochreiter1997long} and GRU~\cite{cho-al-emnlp14} in our setting. RNNs force models to take in tokens sequentially and encode the information recurrently. However, in our context, the main objective is spotting the key entities in texts rather than modelling their semantics. As a results, modeling texts in a recurrent manner might limit the model's flexibility regarding to how they encode relations among different input tokens. Imagine the behavior patterns of human beings. When trying to find matches of text and images, it is likely that we look at both text and image as one piece. We do not try to capture the exact semantic meaning derived from a long and complete sentence but paying attention only to keywords. We further compare these keywords with images and decide if it is a match. 
Attention mechanism also yields more interpretable models. As Self-Attention is designed based on \emph{key}, \emph{value}, \emph{query} and produces attention maps from them, we know exactly what is happening to each word vector and how they are being used to produce features. We discuss this more with examples in Section~\ref{subword-experiments}.

\textbf{\textit{Comparing loss functions.}}
\citet{faghri2018vse++} claimed state-of-the-art text-image retrieval results on MS-COCO~\cite{lin2014microsoft}. However, it performs worse than its prior work~\cite{kiros2014unifying} on our data. The two models mainly differ on the loss function where \cite{kiros2014unifying} adopts a triplet ranking loss that rewards positive pairs and punishes all negative pairs within a mini-batch (same as we do in Equation~\ref{eq:loss}), while \cite{faghri2018vse++}, similarly, also rewards positive pairs, but punishes only the \emph{hardest} negative as shown in Equation~\ref{eq:vsepp-loss}. 
\begin{equation}
\begin{split}
 \mathcal{L_\textsc{Max}} =  \sum_{i\in I} \max_{\bar{t}\in T\backslash \{t\} } [\alpha -s(i,t)+s(i,\bar{t})]_+ + \sum_{t\in T}  \max_{\bar{i} \in I\backslash \{i\}}[\alpha -s(t,i)+s(t,\bar{i})]_+
\end{split}
\label{eq:vsepp-loss}
\end{equation}
In MS-COCO, the descriptions (texts) of an image are usually rich and accurate. The hardest negative within a mini-batch is likely to have a clear difference with the positive. We call them \emph{true} negatives. Put this in another way, the MS-COCO is providing a clear and strong supervision to the model. When it comes to our text-image news dataset, the correspondences among text-image pairs are much weaker. This leads to the fact that, in the context of our weakly supervised data, the hardest negative is likely to contain very little negative information or even contains very positive signals (if it happens to be another descriptive counterpart) - the hard negatives are actually \emph{pseudo} hard negatives. As only the hardest sample is selected, if it happens to be \emph{pseudo}, the gradients of the whole mini-batch would be misleading. In contrast, Equation~\ref{eq:loss} can cancel out false gradients by other correct negatives within the mini-batch, preventing model being harmed by wrong labels. Also, we can tell a bit from the single-textual-source model results in Table~\ref{table:unimodal}. Image caption is the strongest supervision (or most descriptive text) among the four textual sources. It is also the one where the two loss functions' performances have the least difference. It agrees to our interpretation that with a stronger supervision, it is less likely to encounter \emph{pseudo} hard negatives. Despite \emph{pseudo} negatives, noise is also a possible source of error. \cite{wu2017sampling} provides a theoretical analysis of this problem: when negative examples are very hard, the direction of the gradient is likely to be dominated by noise, leading to failure of optimization.

\citet{liu2020hal} recently propose a novel learning objective \textsc{Hal} that makes use of hard samples and avoid \emph{pseudo} hard negatives at the same time. As suggested in Table \ref{table:unimodal}, \textsc{Hal} improves R@10 on caption data by $7-8\%$ without changing any encoder architectures but using the following loss function:
\begin{equation}
\begin{split}
\mathcal{L_\textsc{Hal}} = \frac{1}{N}\sum_{i=1}^{N} \Big( \frac{1}{\alpha}\log(1+\sum_{m\not = i} e^{\alpha (S_{mi}-\epsilon)})
+\frac{1}{\alpha}\log(1+\sum_{n\not=i} e^{\alpha( S_{in}-\epsilon)}) 
- \log (1+ \beta S_{ii} )\Big),
\end{split}
 \label{eq:hal}
\end{equation}
with hyperparameters $N=256,\alpha=20,\beta=30,\epsilon=0.2$. Due to limit of time and space, we leave the experiments combining \textsc{Hal} with our model for future works.

We analyzed the choice of learning objective very carefully as it reveals a gap worth noticing between the datasets used in academia and real-world data. Benchmarks like MS-COCO~\cite{lin2014microsoft}, Flickr8k~\cite{hodosh2013framing} and Flickr30k~\cite{plummer2015flickr30k} are a little bit \emph{too perfect} and may inadvertently harm the generality of models if not used carefully. When models were designed based on certain attributes of the toy datasets, they have potentially followed strong hypothesis which would lead them into overfitting these sets and won't incorporate the real-world data who is usually noisier and weakly labeled.

\subsubsection{Multi-Head Attention as a query system}
\label{query-system}

\begin{SCfigure}
\vspace{-0.3cm}
\centering
\includegraphics[width=8.5cm]{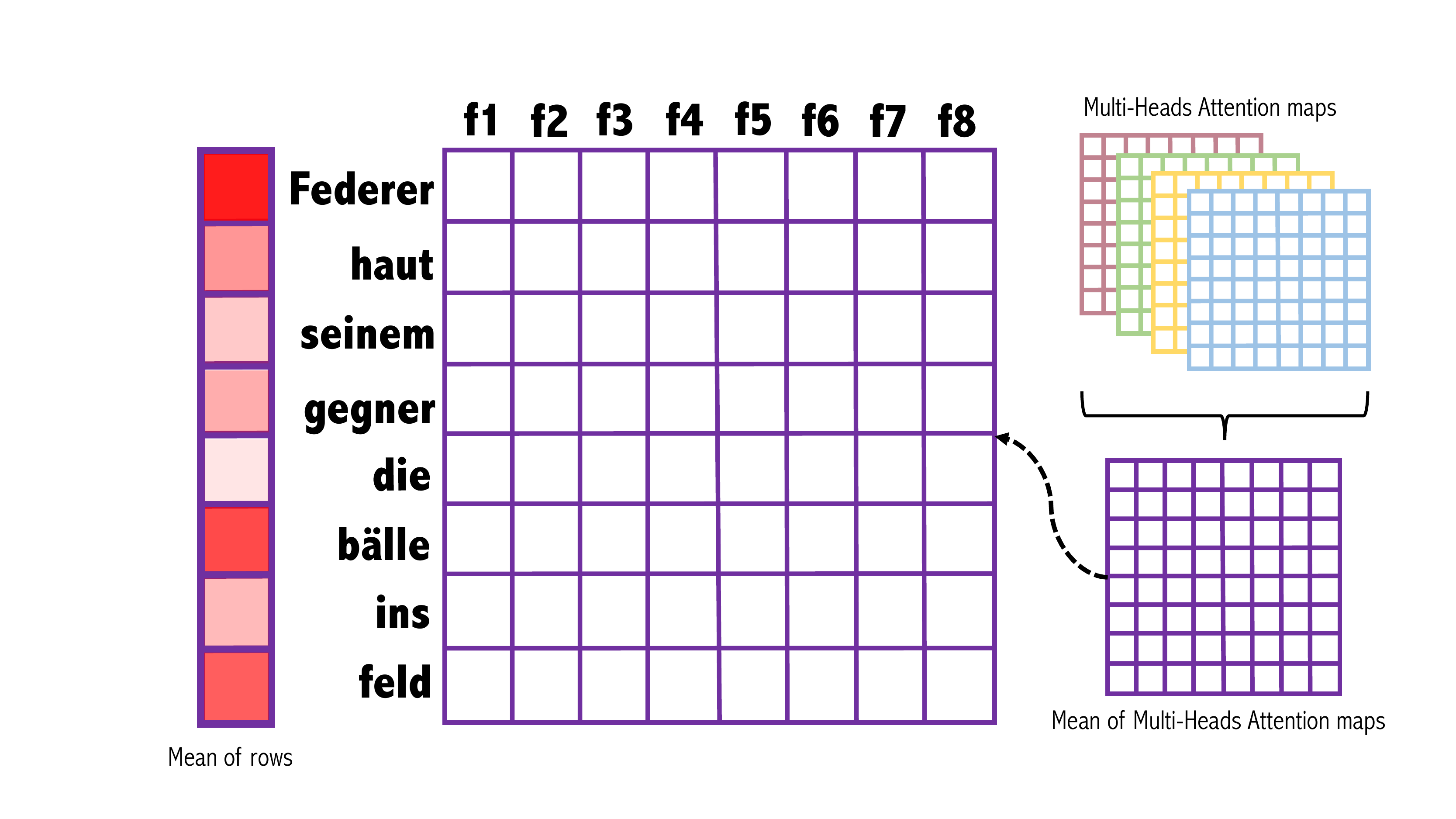}
\caption{From right to left: compute the mean of feature maps and compute the mean of each row. Attention maps are obtained from $M$ computed by Equation~\ref{eq:attn-3}.}
\label{fig:attn-score-explained}
\vspace{-0.3cm}
\end{SCfigure}

We argue that the Multi-Head Attention mechanism is acting as a query system to filter out unimportant words (information) and preserve the important ones. We validate this by visualizing Multi-Head Attention layer's attention maps. We do this by averaging rows of averaged Multi-Head attention maps. 

A toy example of explaining how to compute attention scores can be found in Figure~\ref{fig:attn-score-explained}. The columns denote features, which are the outputs of the Multi-Head Attention layer; while the rows denote embeddings of words, which are the inputs of the Multi-Head Attention layer. For an $h$-Head Attention, it produces $h$ attention maps of size $(l,l)$ where $l$ is the length of the input sequence. We average these $h$ maps to obtain a single attention map as suggested in the figure. For an output feature $i\in \{1,\cdots,l\}$, githuthe $i$-th column of attention map specifies weights assigned to each input word. In other words, the $i$-th column can be interpreted as the importance of each word to feature $i$. And accordingly, the $j$-th row ($j\in \{1,\cdots,l\}$) can be interpreted as how important word $j$ is to all the output features. The more a word contributes to (all) output features, the higher the weights would be on that row. Then it makes sense to average rows. Words being paid more attention would have higher average scores.

\begin{figure}[h]
\vspace{-0.1cm}
\centering
\includegraphics[width=14cm]{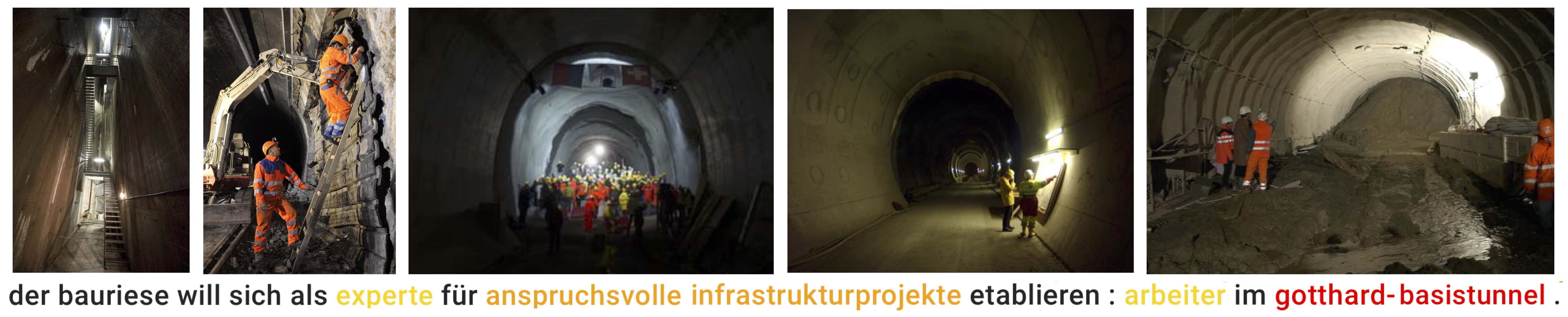}
\includegraphics[width=14cm]{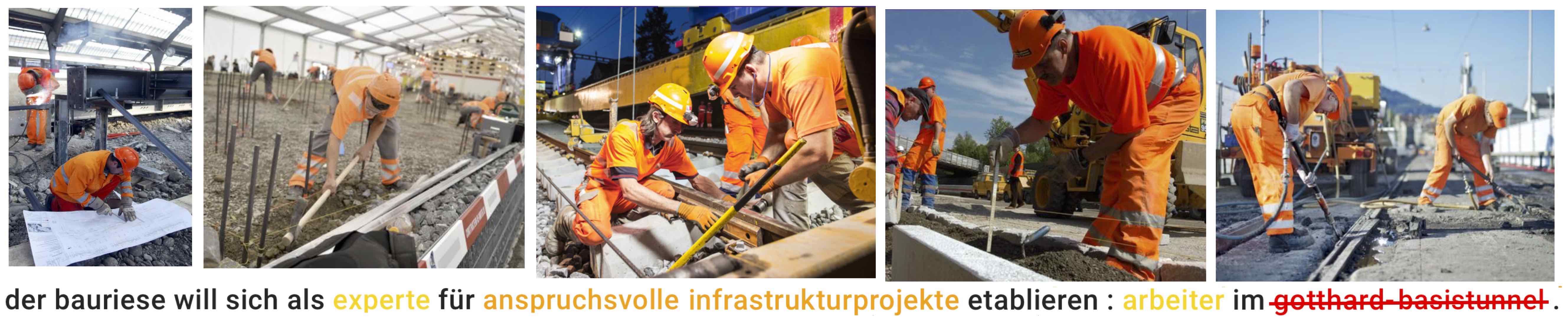}
\includegraphics[width=14cm]{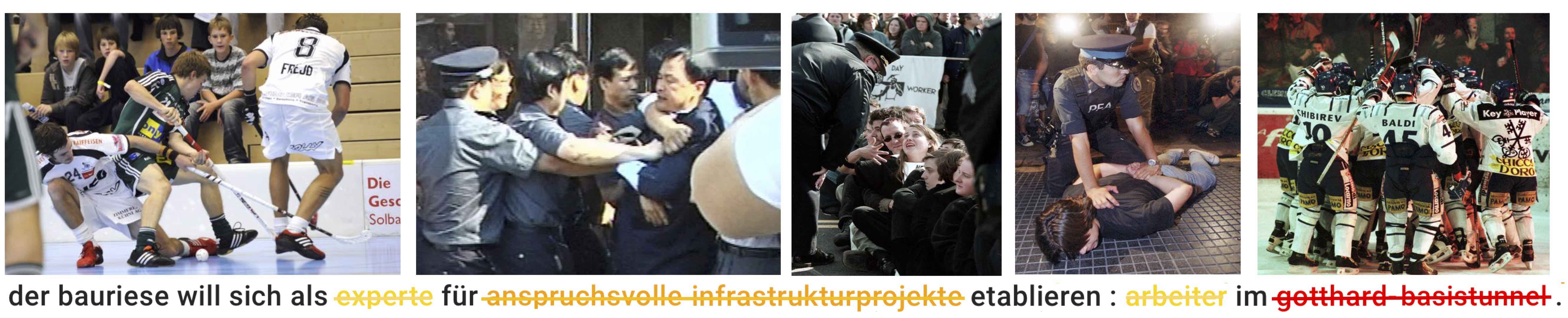}
\includegraphics[width=14cm]{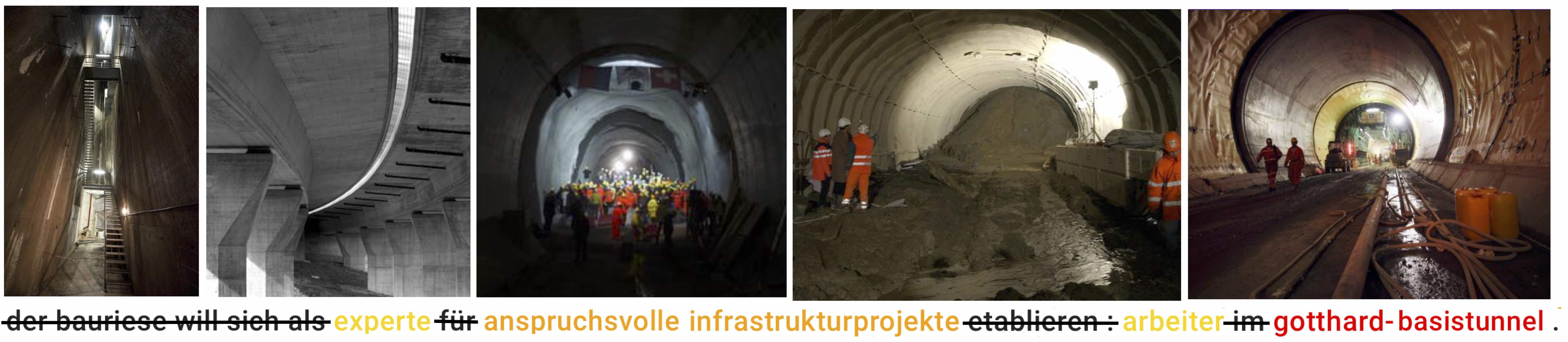}
\caption{The first row is searching with full text. The key components \emph{Basistunnel} (base tunnel) and \emph{arbeiter} (workers) are both shown in most of the top 5 retrievals. The second row is searching without the token in red. \emph{arbeiter} are still presented however the \emph{Basistunnel} is gone and scenes become very random. The third row is searching without all highlighted words. The results become completely irrelevant. The fourth row, on the contrary, is searching with only the highlighted words. Retrievals are comparable to searching with full text. (Best seen in colors.)}
\label{fig:query-system-explained}
\vspace{-0.3cm}
\end{figure}

Inside the text encoder, Multi-head Attention and Point-Wise CNN are the only two learnable parts. As all kernels of Point-Wise CNN are of size $1$, they do not perform any \emph{weighted sum} operation and treat all vectors in the input matrix equally. So, Multi-Head Attention is mainly responsible for querying out useless information as it is doing a global \emph{weighted sum} and the network has sufficient freedom to drop worthless information with simply changing $W^Q$ and $W^K$ in query-key weight matrix $QK^T = (EW^Q)(EW^K)^T$ (see Equation~\ref{eq:attn}).

As we argue that attention is important for our task, we do an extensive analysis on how it works in the context of retrieving relevant news images.
We play with the following toy example:

\begin{compactitem}
    \item \emph{Der Bauriese will sich als experte f\"ur anspruchsvolle infrastrukturprojekte etablieren: arbeiter im Gottard-Basistunnel. (20. November 2014)} \footnote{\emph{The construction giant wants to establish itself as an expert for sophisticated infrastructure projects: workers in the Gottard Base Tunnel. (November 20, 2014)} (English translation)}
\end{compactitem}

Searching this text with our model trained only with captions, the visualization of attention scores received are presented in the first row of Figure~\ref{fig:query-system-explained} where attention decreases in the order of red, orange, yellow and black. We compare the qualitative results of searching with: i) full text; ii) the token with the highest score crossed out\footnote{When crossing out tokens, a meaningless out-of-vocab placeholder would be used to replace the original ones. We used a random emoji to do the job.}; iii) all highlighted tokens crossed out; iv) only the highlighted words. All the results can be found in Figure~\ref{fig:query-system-explained}.

We can clearly see that searching with the highlighted words is comparable to searching with full text. On the contrary, the removal of highlighted words decreases the quality of retrieved images.

\subsubsection{Subword information matters (a lot)} \label{subword-experiments}
In our context of retrieving images using short texts (like headline, lead, caption), there might only be a few informative words within one piece of text. These words are usually entities (as news is about entities). If these entity words are missed, the retrieval task will almost certainly fail. We argue that our word\&subword embedding layer handles entity names by 1) tolerating typos; 2) recognizing compound and conjunction words and 3) transferring knowledge across close languages. Please refer to Appendix for specific examples.

\begin{table*}[h]
\small
\renewcommand{\arraystretch}{0.9}
\setlength{\tabcolsep}{3pt}
\caption{Comparing different multi-textual-source feature fusion methods.}
\label{table:1}
\centering
\begin{tabular}{clcccccccc}
\toprule
\multirow{2}{*}{\#} & \multirow{2}{*}{\bf Fusion Method}  & \multicolumn{4}{c}{\bf image to text} & \multicolumn{4}{c}{\bf text to image} \\ \cmidrule(l){3-6} \cmidrule(l){7-10}
  & &  R@1 & R@5 & R@10 & Med r & R@1 & R@5 & R@10 & Med r \\ \midrule
2.1 & Global Max Pooling &  21.4 & 50.0 & 64.3 & 5.0  & 19.6  & 47.4 & 62.3 & 6.0   \\
2.2 & Global Max Pooling (w/o random drop) &  22.0  &  50.3 &  65.2  &  5.0 & 17.9 & 47.6 &   \textbf{63.5} & 6.0  \\
2.3 & Element-wise Adding & 21.2 & 50.6 & 64.4 & 5.0   &\textbf{20.8} & \textbf{49.4} & 63.1 & 6.0   \\
2.4 & Element-wise Adding (w/o random drop) & 21.8           & 50.5    & 65.6  & 5.0 & 17.2  & 46.3  & 62.0 & 6.0  \\
2.5 & Neural Net Fuser &  24.7 & 53.2 & 68.7 & 5.0  & 17.8 & 47.8 & 63.0 & 6.0   \\
2.6 & Neural Net Fuser (w/o random drop)  & \textbf{25.6}  & 54.0  & 67.8 & 5.0 & 20.2 & \textbf{49.4}   &  62.7   & 6.0 \\
2.7 & Attention Fuser (from scratch) &  14.9 & 39.4 & 54.3  & 9.0   & 12.1 &    33.2 & 47.9 & 12.0\\
2.8 & Attention Fuser (w/o random drop)  & 22.8 & 51.7 &   66.6 & 5.0 & 19.3   &  47.2 & 62.8 & 6.0  \\
2.9 & Attention Fuser & 24.3 & \textbf{54.8} & \textbf{70.6} & 5.0   & 18.2 & 46.8 & 62.7 & 6.0  
\\ \bottomrule
\end{tabular}
\vspace{-0.2cm}
\end{table*}

\subsection{Fusion of Multiple Textual Sources}
\label{sec:experiment-multi-text}

Preserving everything we had for single textual source, in this section we further experiment methods for fusing information obtained from different textual inputs. We start by introducing two techniques used during training - \emph{Divide-and-Conquer} and \emph{random drop}. Then we compare different fusion policies including both pre-defined and learnable ones.

\textbf{\textit{Divide-and-Conquer: Starting with Single Source.}} 
We find that the multi-textual-source model can greatly benefit from starting with weights trained on single textual source. Specifically, for each encoder in the \emph{fourway} model, we load their corresponding single-textual-source encoder's weights. Since the four single models share a common word embedding in the \emph{fourway} model, we initialize the word embedding layer with caption model's word embeddings weights (which is the one with the best single-textual-source performance). To avoid these pre-trained weights receiving random gradients at the starting phase of joint training, we freeze them for $1$ epoch during which we only train the \emph{Feature Fuser}. Then all parts of text encoder are unfrozen and being trained as a whole for another $9$ epochs. We use a learning rate of $10^{-4}$ for the first $5$ epochs and $2\times10^{-5}$ for the second $5$ epochs. We empirically find it very important to start with what we already have as suggested in Table~\ref{table:1}. R@10s are improved by $\approx \textbf{15\%}$! Training four text encoders together from scratch might be too hard a task to optimize.

\textbf{\textit{Random Drop.}} 
To tackle possible missing sources of text inputs in real-world applications, we use a very simple \emph{random\_drop} mechanism and it unexpectedly improves R@10s of models with learnable fusion policy.

Briefly, texts from each sources have a probability of being dropped and substituted with empty string during the training phase. For every sample, one source of texts is randomly chosen to be preserved (to make sure at least one meaningful source of input) while for the other three sources, each one has $0.7$ probability of being kept and $0.3$ probability of being dropped. We switch back to \emph{complete\_inputs} mode in the validation phase. According to the validation results in Table~\ref{table:1}, models with learnable fusion policy (\emph{Neural Net Fuser} and \emph{Attention Fuser}) using \emph{random\_drop} is doing even better than using \emph{complete\_inputs} in most indicators. 

We infer that \emph{random\_drop} is acting as a regularizer in our model. The deep learning community has developed robust and effective regularization techniques like dropout~\cite{srivastava2014dropout} which randomly suppresses neurons of the network by simply erasing their outputs; or ShakeDrop~\cite{yamada2018shakedrop}, which even reverses signs of them. Also, in computer vision, it is very common to apply random crop on original images to alleviate model from overfitting. Very recently, \cite{devries2017improved,zhong2017random} propose \emph{random erasing data augmentation} which is to randomly erase patches of input images (and fill them with zeros or random values). These are also regarded as  regularization methods. \emph{random\_drop} is following the same idea of randomly hiding part of the input to prevent the model from relying on some specific feature (or textual source, in our context).

\textbf{\textit{Other Fusion Methods.}} 
We also test other fusion methods in the literature such as Global Max Pooling, Element-wise Adding, Neural Nets Fusion. We use similar manner loading pre-trained single-textual-source models, then train them jointly with random drop. Global Max Pooing is taking the max value of the four outputs for every element; Element-wise Adding means adding features from every textual source directly, which should be working similarly as Global Average Pooling; Neural Nets Fusion comes from \cite{zahavy2016picture} where they find stacking two blocks of (fully-connected layer + ReLU activation) is the optimal structure for their multimodal feature fusion task. We compare these results with our Attention Fuser in Table~\ref{table:1}. We notice that Element-wise Adding performs better in image-to-text task and Attention Fuser is better for text-to-image.

\subsection{Multilingual Model}
\label{sec:exp-multilingual}



We adopt the recently released \textbf{M}ultilingual \textbf{U}nsupervised or \textbf{S}upervised word \textbf{E}mbeddings (\emph{MUSE}). Comparing to fastText \emph{wiki} which we used in previous experiments, \emph{MUSE} is rather new and has a relatively small vocabulary. Since it is being adjusted for alignment, the quality of its numerical representations may also be compromised. However, it is promising as a solid step towards erasing the language barrier in NLP and machine learning. Experiments show that by using \emph{MUSE} to bridge multilingual data, our multilingual model can already beat the best single-textual-source model without subword information (previous best: $48.1\%$, \emph{MUSE} best: $48.5\%$). It is, however, less powerful than the best model with subword embeddings. We expect better multilingual word embeddings considering subword information emerge in the near future. 

We compare a single-textual-source bidirectional model being trained on monolingual data and trained on bilingual data (both on image captions), following the process described in Section~\ref{sec:14all}. As suggested in Table~\ref{table:one-for-all}, the language-agnostic model performs better than the monolingual model in both German and French.

\begin{table}[h]
\small
\renewcommand{\arraystretch}{0.9}
\caption{Quantitative results of one-for-all. R@10 is reported here.}
\label{table:one-for-all}
\centering
\begin{tabular}{clcccccc}
\toprule
\multirow{2}{*}{\parbox{0.2cm}{\#}} & \multirow{2}{*}{\parbox{1.4 cm}{\bf training language}} & \multicolumn{2}{c}{\bf image to text}   & \multicolumn{2}{c}{\bf text to image} \\ \cmidrule(l){3-4} \cmidrule(l){5-6}
 & & de & fr &  de & fr \\ \midrule
3.1 & fr & - & 39.4 & - & 38.2 \\
3.2 & de & 41.1 & - & 41.4 & - \\
3.3 & de + fr & \textbf{43.1} (+2.0) & \textbf{42.0} (+2.6) & \textbf{43.3} (+1.9) & \textbf{41.7} (+3.5) \\
\bottomrule
\end{tabular}
\end{table}

Notice that the German single-textual-source model (with regular word embeddings) results in Table~\ref{table:unimodal}, Table~\ref{table:one-for-all} and Table~\ref{table:all-for-one} are all different (\emph{line 1.10, 3.2, 4.2}). In terms of R@10 for text-to-image, they are \emph{line 1.10}: $45.8\%$, \emph{line 3.2}: $41.1\%$ and \emph{line 4.2}: $42.1\%$ respectively. This is due to the three experiments differ in vocabulary and choice of pre-trained vectors. All three are using only regular full word embeddings pre-trained on Wikipedia and without subword information, however, \emph{1.10} is using \emph{wiki}~\cite{bojanowski2017enriching} and the word embedding is finetuned while training while \emph{3.2} and \emph{4.2} are using \emph{MUSE}~\cite{lample2017unsupervised}. Though also pre-trained on Wikipedia, \emph{MUSE} is further processed and aligned with word vectors of other languages. And more importantly, to better align with other languages, \emph{MUSE} only contains 200k word vectors while \emph{wiki} has over 2275k, which is more than 10 times larger. Using \emph{MUSE} leads to many more out-of-vocab words who might be the main cause for the drop in performance. Though both using \emph{MUSE}, \emph{line 3.2} and \emph{line 4.2} differs in vocabulary and training. To be consistent with the \emph{one-for-all} experiment setting, \emph{line 3.2} uses a frozen word embedding layer to prevent the aligned multilingual space from vanishing while \emph{line 4.2} finetunes its word embedding. As \emph{line 3.2}'s word embeddings are frozen, all out-of-\emph{MUSE}-vocab words are considered as unknown while in \emph{line 4.2} learns out-of-\emph{MUSE}-vocab word vectors on the fly. This also explains the difference in French baseline results in Table~\ref{table:one-for-all} and Table~\ref{table:all-for-one} (\emph{line 3.1, 4.1}).


Also, in the \emph{all-for-one} setting as described in Section~\ref{sec:all41}, where maintaining aligned word embedding is no longer required, the model trained on both German and French has improved $7.7\%$ in German and $4.1\%$ in French for R@10 scores in text-to-image retrieval as suggested in Table~\ref{table:all-for-one}.

\begin{table}[h]
\small
\renewcommand{\arraystretch}{0.9}
\caption{Quantitative results of all-for-one. R@10 is reported here.}
\label{table:all-for-one}
\centering
\begin{tabular}{clcc}
\toprule
\# & {\bf training language} & {\bf image to text} & {\bf text to image} \\ \midrule
4.1 & fr, full vocab                 & 41.0                 & 39.9 \\
4.2 & de, full vocab               & 42.1                 &
42.2 \\
\midrule
4.3 & de (+ fr), \emph{MUSE} vocab         & 46.4  (+4.3)               & 44.1 (+1.9)             \\
4.4 & de (+ fr), full vocab       & \textbf{49.8} (+7.7) & \textbf{48.5} (+6.3) \\
4.5 & fr (+ de), \emph{MUSE} vocab     & 44.9 (+3.9)          & 43.2 (+3.3)          \\
4.6 & fr (+ de), full vocab         & \textbf{45.1} (+4.1) & \textbf{43.4} (+3.5) \\
\bottomrule
\end{tabular}
\end{table}

Notice that it is also very important to add out-of-\emph{MUSE}-vocab vocabularies into word embedding. As these words might contain lots of language-specific knowledge. It is especially important for languages with very large vocabularies (like German). As suggested in Table~\ref{table:all-for-one}, R@10 only improves $4.3\%$ if not adding full vocabulary while there is $7.7\%$ improvement with full vocabulary.

%% file: contents/limitation.tex
\section{Limitation}
\label{sec:limitation}
In this section, we analyze the major limitations of our model and how we could fix some of them. 

\emph{\textbf{What's the Real Ground-truth?}} What should be considered a \emph{correct} image for a news article is subjective, and the standards are heterogeneous. The ambiguity of the real article-image relationship has caused inconvenience in both training and evaluating models. As described in the weakly-supervised learning setting, the model learns from article-image pairs with very different levels of correspondences during training, and some samples might even be contradicting. During the evaluation, the quantitative metrics only roughly measure the correspondence of articles and images. However, it remains unclear how exactly judgments should be made towards \emph{correct} retrievals.

\emph{\textbf{Photo Illustrations and Factual Photos.}} Our model works the best on retrieving \emph{photo illustrations}\footnote{Photo Illustrations, or Photo Pr\'etexte in French, is a notion from Journalism that refers to non-factual photos that visually illustrate contents of news articles. They usually come from one of the image banks to which most media subscribe or each media's own image collection. It is a controversial topic in the news industry on how photo illustrations should be used, but they are only getting more and more popular on the web today as ``an article without an image has no chance of generating traffic''~\cite{foto-pretext}.} rather than factual photos with specific named entities in it. Photo illustrations mainly appear in news articles relating to general but random topics such as car accident, traffic jam, snowstorm, pregnancy, university lab, school, etc.. The retrieval quality for named entities, however, heavily relies on the frequency that named entity appears in the (training) database. For instance, it may perform well on really famous people, eg. ``Donald Trump'', but may miss less notable ones. There are also classes of entities having only subtle visual differences. For instance, when the query is ``Lake Geneva'', the model also retrieves many other lakes as they look alike. One possible improvement can be creating more sophisticated visual encoders. For instance, finetuning image encoders, adding object detection modules and transfer learning from a specific domain could be plausible options.

\emph{\textbf{A Fix - Named Entity Filtering.} }
News images are usually taken and stored in a database with \emph{metadata}. The metadata is either given by the photographer or editors and usually contain the main named entity presented in the image. Traditionally, news media use keyword matching to search for related images (text-to-text search). As long as the metadata is accurate, retrieving the correct named entities is guaranteed. We can partially fix the model's weakness in retrieving named entity photos by matching the retrievals' metadata with named entities detected in the news articles. Specifically, we use \cite{spacy2} to detect four sorts of named entities - PER, LOC, ORG and MISC that are presented in the news article. Then a user may select the desired entity(ies) and all photo candidates without that entity in its metadata would be filtered out. Notice that the combination of our model and metadata search is more powerful than metadata search only as it puts entities in context (sentiment, gestures, companions, environments, etc.). We provide some more specific examples in the Appendices.

%% file: contents/conclusion.tex
\section{Conclusion and Future Work}
\label{sec:conclusion}

Our work proposes the News Image Selection task in the context of weakly-supervised learning and large-scale database. We introduce an automated image selection system to assist photo editors in a multilingual multimodal setting. Our system incorporates latest advancement in neural machine translation and deep learning community like subword embedding and self-attention mechanism. We also propose novel techniques for multimodal and multilingual learning that contributes to the natural language understanding and machine learning literature. We also extensively discuss problematic issues in News Image Selection, explaining how it has made state-of-the-art text-image retrieval method failed on our dataset. And we analyze the limitations of our proposed method and show how combining traditional metadata search can help mitigate some of the problems. Besides News Image Selection, the system can potentially be used for other text-image recommendation scenarios. As this work mainly focuses on enhancing text encoder, we plan to improve image encoder as well in future works, e.g. incorporating a pre-trained object detector \cite{ren2015faster} to enhance image encodings. We also plan to explore improve the textual encodings by exploring more variations of recently emerged pre-trained/pre-aligned multimodal/multilingual embeddings \cite{devlin2019bert,chen2019unsupervised,su2020vlbert}.

%% file: contents/ack.tex
\section{Acknowledgement}
We thank Dr Hamza Harkous for helpful discussions and feedbacks. We thank anonymous reviewers for their comments based on which we were able to enhance our work. We gratefully acknowledge the support of NVIDIA Corporation with the donation of the Titan Xp GPU used for this research.

%% file: contents/appendix.tex
\section{Appendices}

\subsection{Impact of modeling word on the char level}

Opting for a subword embedding layer enables three features for our system: 1) tolerating typos; 2) recognizing compound and conjunction words and 3) transferring knowledge across close languages. In the following we show real examples and retrieval results for demonstration.

\begin{figure}[h]
\centering
\includegraphics[width=8cm]{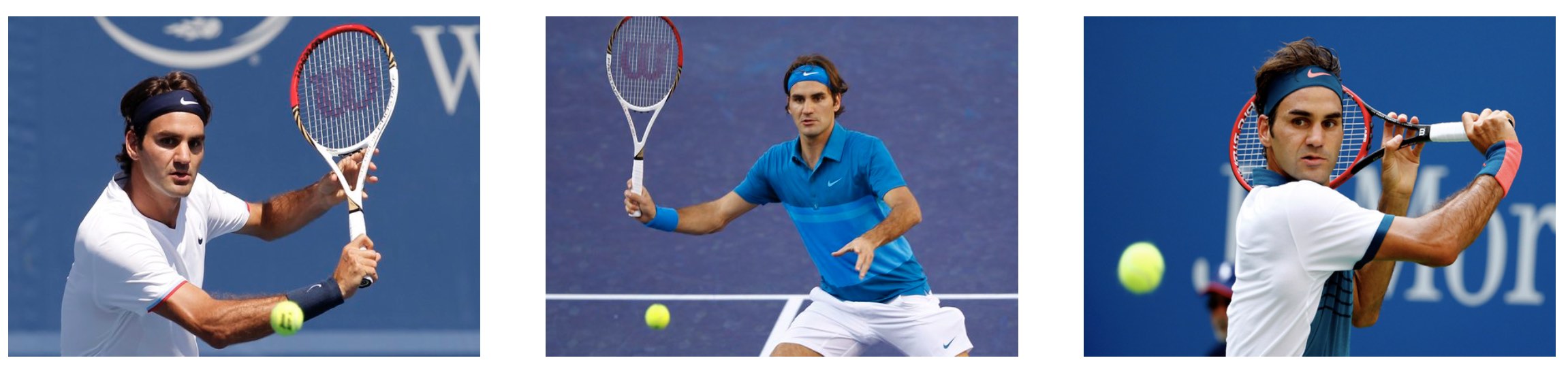}
\includegraphics[width=8cm]{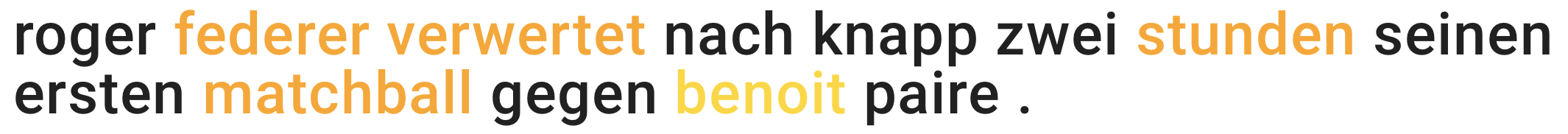}
\includegraphics[width=8cm]{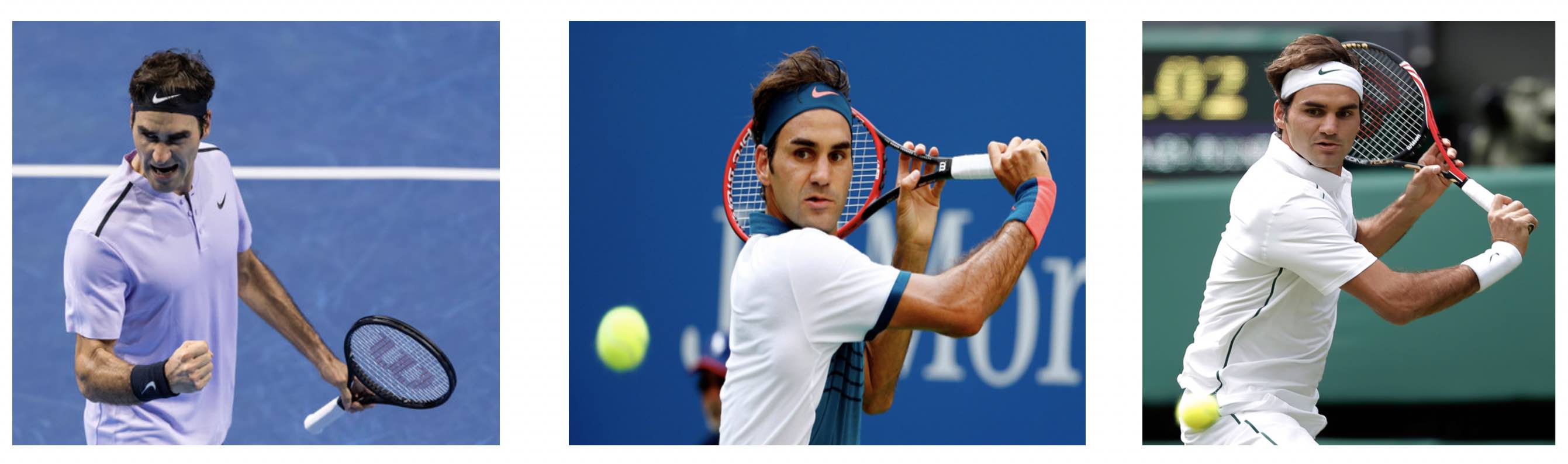}
\includegraphics[width=8cm]{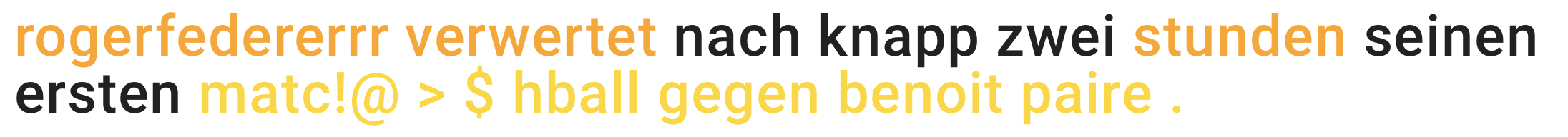}
\caption{Top 3 retrieval results of the original sentence (first row) and with typos (second row). The two rows are of the same quality. (Best seen in colors.)}
\label{fig:typos}
\end{figure}
\textbf{\textit{Tolerate Typos.}}
Subword embedding can tolerate typos rather than regarding everything rare as out-of-vocab words. This is extremely important in real-world applications. 
\begin{compactitem}
    \item \emph{{\color{Red}Roger Federer} verwertet nach knapp zwei Stunden seinen ersten {\color{Blue}Matchball} gegen Benoit Paire.}\footnote{\emph{Roger Federer exploits his first match ball against Benoit Paire in less than two hours.} (English translation)}
    \item \emph{{\color{Red}RogerFedererrr} verwertet nach knapp zwei Stunden seinen ersten {\color{Blue}Matc$!@>\$$hball} gegen Benoit Paire.}
\end{compactitem}
The first is the original sentence and the second is with typos where most informative words \emph{{\color{Red}Roger Federer}} and \emph{{\color{Blue}Matchball}} are wrongly spelled and unlikely to be in the vocabulary. However, subwords of the typos would still contain enough information to retrieve the correct results as suggested in Figure~\ref{fig:typos}.

\textbf{\textit{Compound and Conjunction Words.}} 
Let's look at two examples. The first example is for compound words. In this news~\cite{derbund-apartments} which talks about building new apartments in the city of Bern, the below three words are the only informative ones in image caption:

\begin{compactitem}
\item \emph{Miet{\color{Red}{wohnungen}}} (rental {\color{Blue}apartment}s)
\item \emph{Stockwerkeigentums{\color{Red}{wohnungen}}} ({\color{Blue}condominium}s)
\item \emph{Alters{\color{Red}{wohnungen}}} (senior {\color{Blue}apartment}s)
\end{compactitem}
 The subword \emph{{\color{Red}{wohnungen}}} means {\color{Blue}apartment} in English, the compound forms are however not likely to be in the vocabulary. With subword information, searching with this kind of text would still work.
 
The second example is for conjunction words. Again, the following word would be the only informative word in the headline of this news~\cite{easyjet-flug} discussing low-cost airlines.
\begin{compactitem}
\item \emph{{\color{Red}easyjet}-{\color{Blue}flug}} ({\color{Red}easyjet} {\color{Blue}flight})
\end{compactitem}

\emph{{\color{Red}easyjet}-{\color{Blue}flug}} is likely to be identified as unknown words in regular word embeddings.  With n-gram subword embeddings, subwords {\color{Red}\emph{easyjet}} and {\color{Blue}\emph{flug}} would both be kept as part of the final representation of \emph{{\color{Red}easyjet}-{\color{Blue}flug}}.

\textbf{\textit{Transferring knowledge across nearby languages.}}
Names of entities are usually very similar in terms of morphology in nearby languages.

\begin{figure}[h]
\centering
\includegraphics[width=9.0cm]{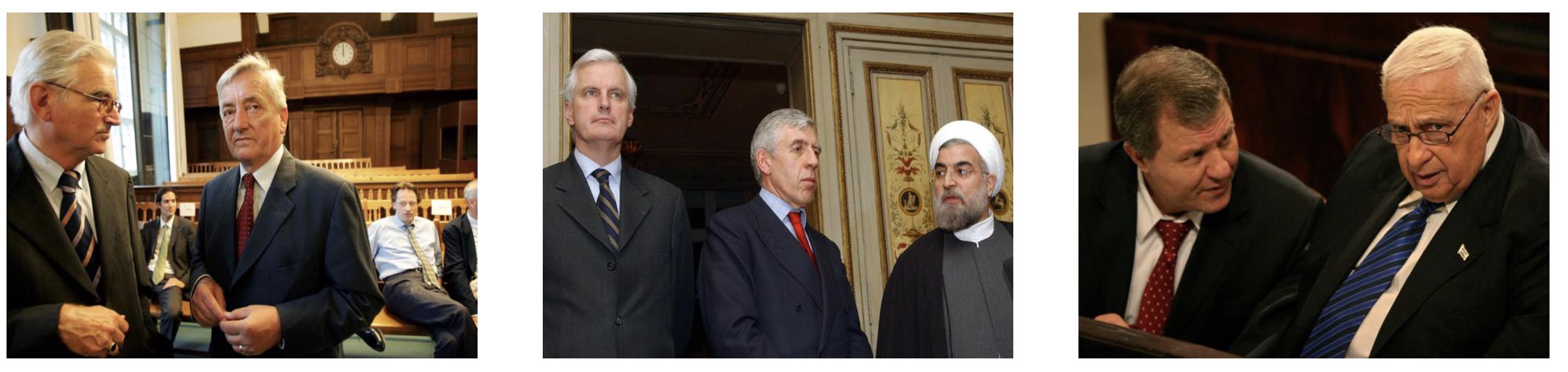}\\
\includegraphics[width=7.0cm]{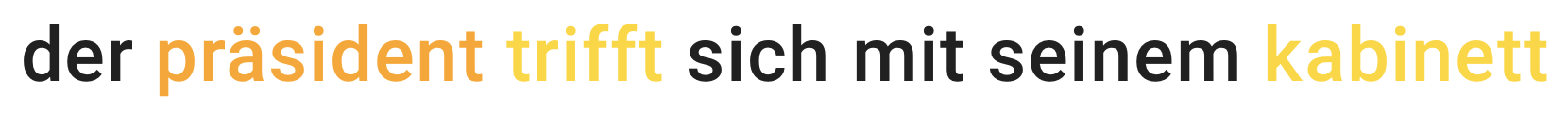}\\
\includegraphics[width=8.7cm]{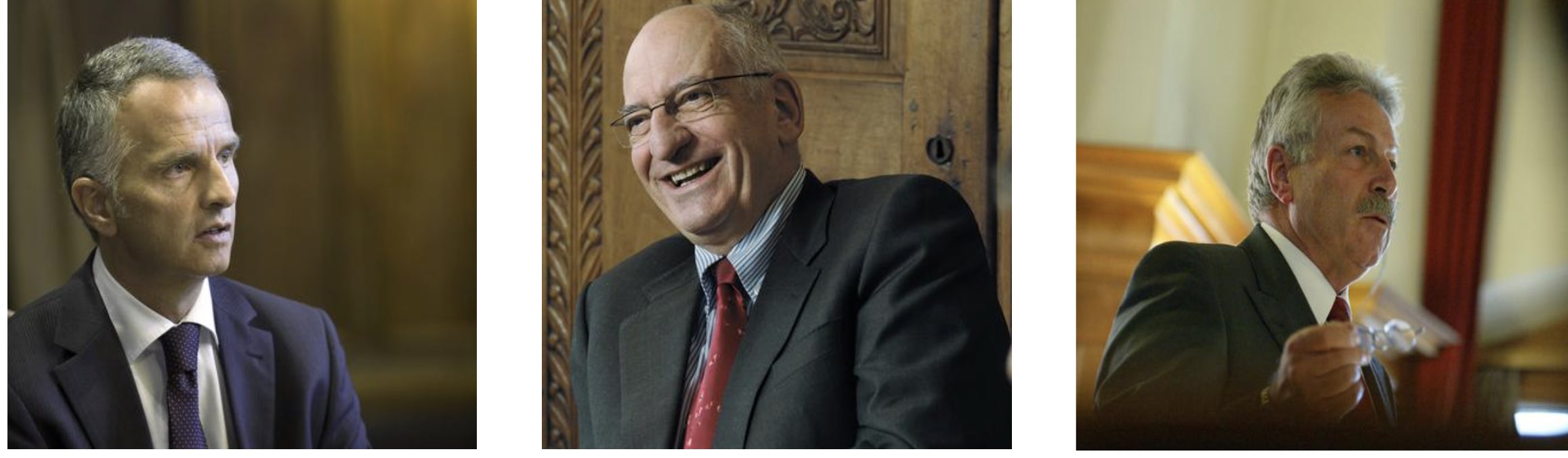}\\
\includegraphics[width=5.5cm]{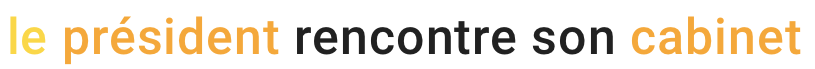}\\
\includegraphics[width=9.0cm]{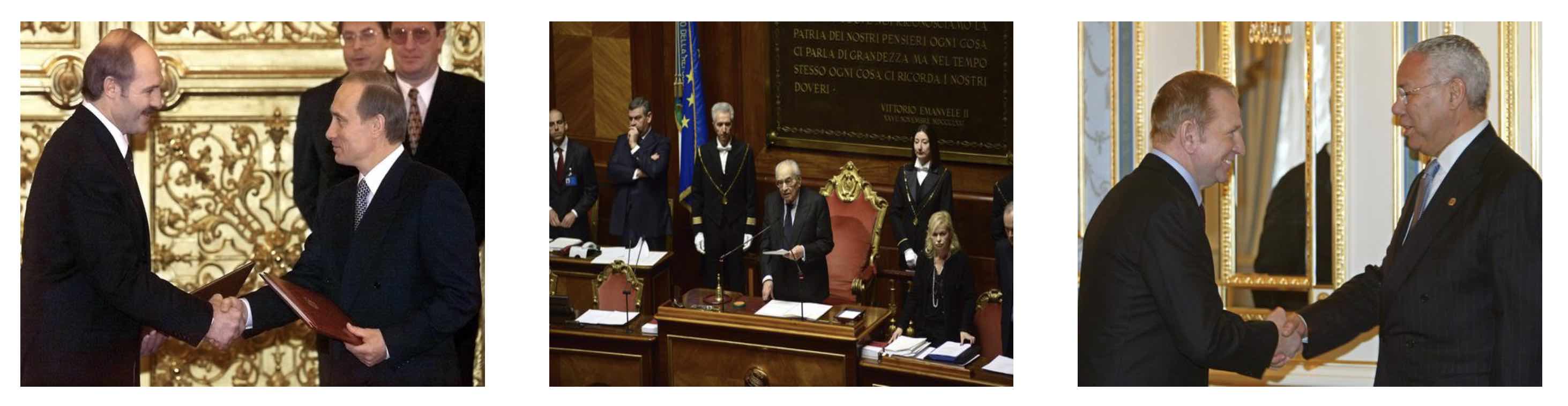}\\
\includegraphics[width=6.5cm]{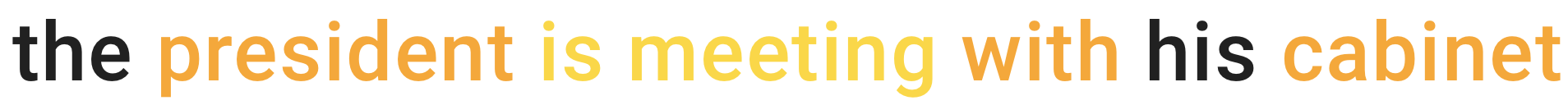}
\caption{The three rows are the top 3 retrieval results of searching in three different languages respectively. The highlighted words from Multi-Head Attention also suggests that \emph{{\color{Red}President}} and \emph{{\color{Blue}cabinet}} received the highest attention scores in both three languages. (Best seen in colors.)}
\label{fig:presidents}
\end{figure}

Let's look at an example of three different \emph{Presidents}:
\begin{compactitem}
    \item \emph{Der {\color{Red}Pr\"esident} trifft sich mit seinem {\color{Blue}Kabinett}.} (German)
    \item \emph{Le {\color{Red}Pr\'esident} rencontre son {\color{Blue}cabinet}.} (French)
    \item \emph{The {\color{Red}President} is meeting with his {\color{Blue}cabinet}.} (English)
\end{compactitem}
The two informative words \emph{{\color{Red}President}} and \emph{{\color{Blue}cabinet}} have many n-grams overlaps among all three languages. We train the caption model only on German and test it also on French and English. Though the model has never seen French and English before, the qualitative results still look relevant as shown in Figure~\ref{fig:presidents}.

\subsection{Combine our system with metadata matching}

We show two examples of combining our model with metadata matching. It aims to demonstrate how the model performs when searching for non-illustrative ``factual'' photos.

\subsubsection{Le Pr\'esident}
\begin{figure}[h]
\centering
\includegraphics[height=6.0cm]{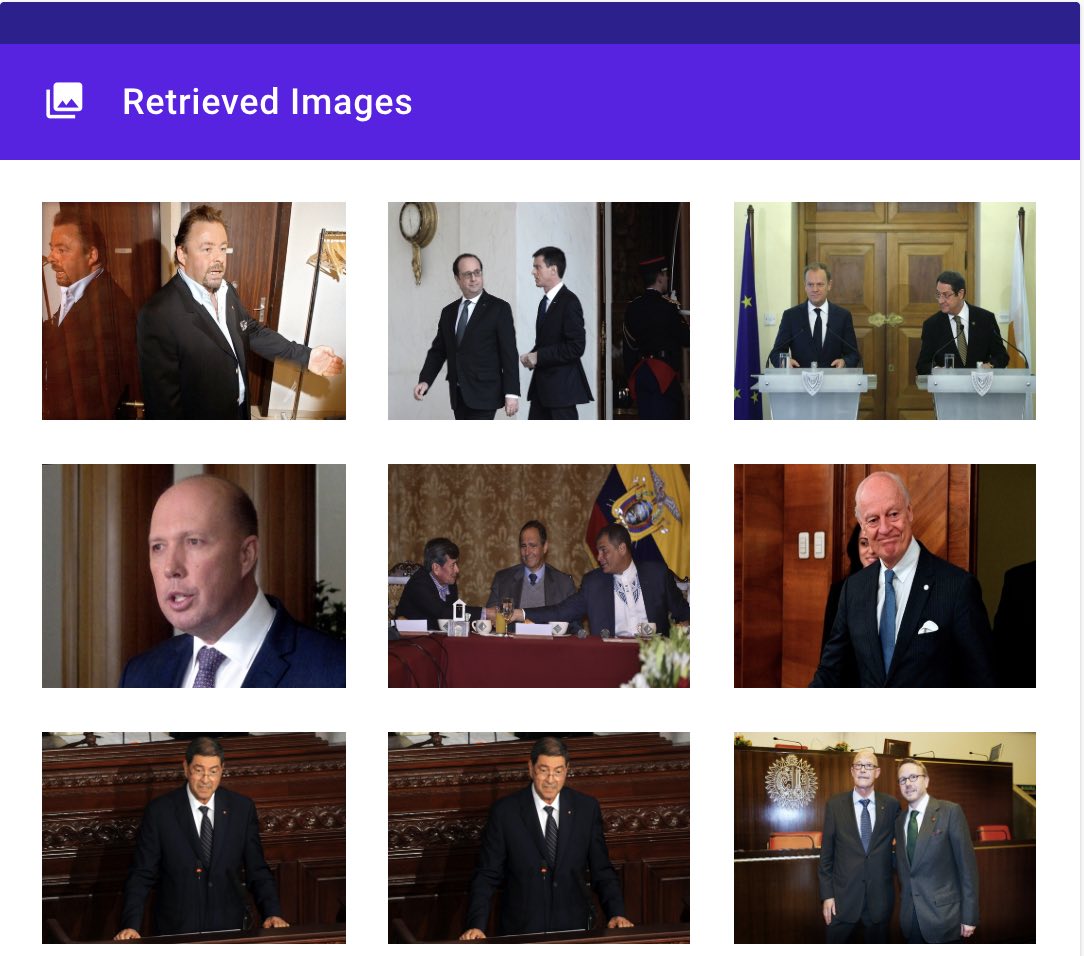}
\includegraphics[height=6.0cm]{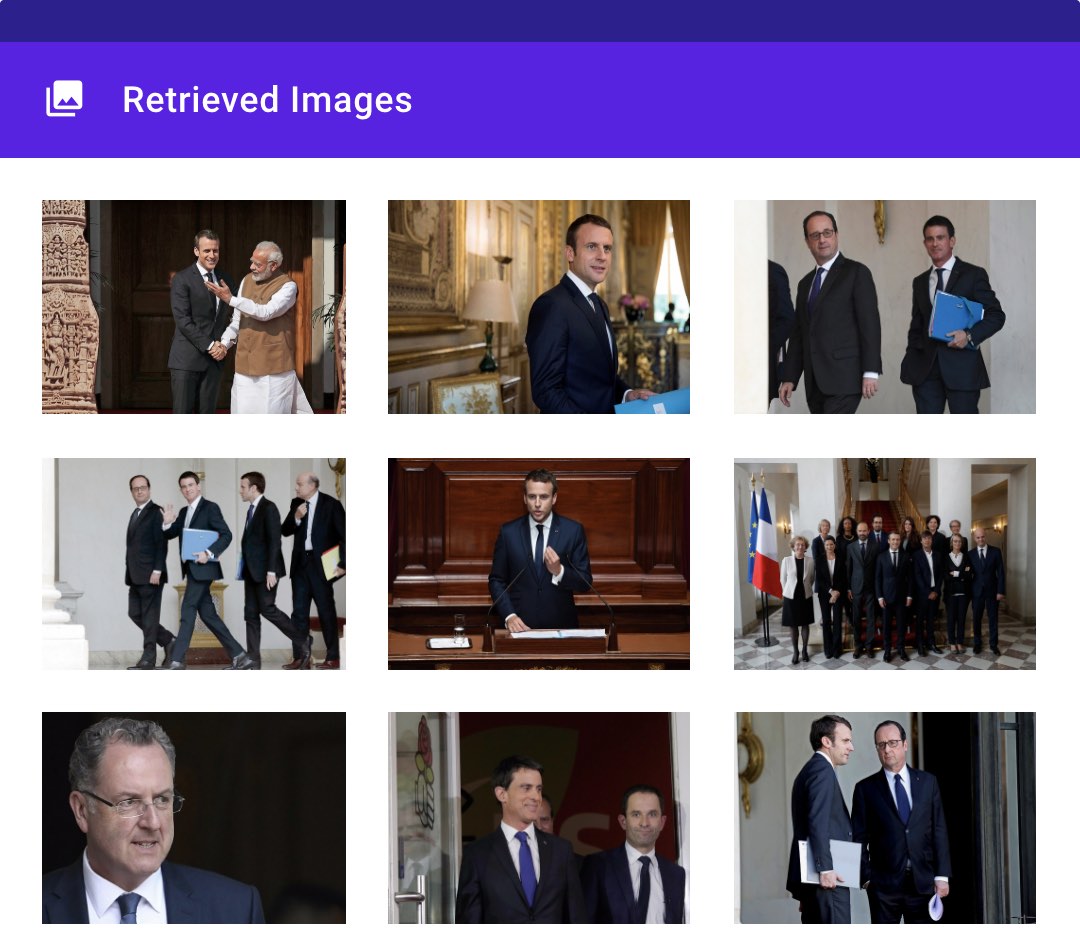}
\caption{Top 9 retrieval results of the query ``\emph{Le Pr\'esident rencontre son cabinet.}'' as title. The left is without selecting any named entity; the right is searching with named entity ``Macron''.}
\label{fig:macron}
\end{figure}

Figure~\ref{fig:macron} shows the top 9 retrievals with the French query ``\emph{Le Pr\'esident rencontre son cabinet.}''\footnote{English translation: \emph{The President is meeting with his cabinet.}} as a title. The left figure is searching without selecting any named entity. The result seems to be any politician(s) meeting or speaking as it's unclear who the ``Pr\'esident'' refers to. The right figure is searching with named entity ``macron'', then 7 of the 9 retrievals contain the French President Emmanuel Macron meeting people or speaking. 

\subsubsection{Der Z\"urichsee}
\begin{figure}[h]
\centering
\includegraphics[height=6.0cm]{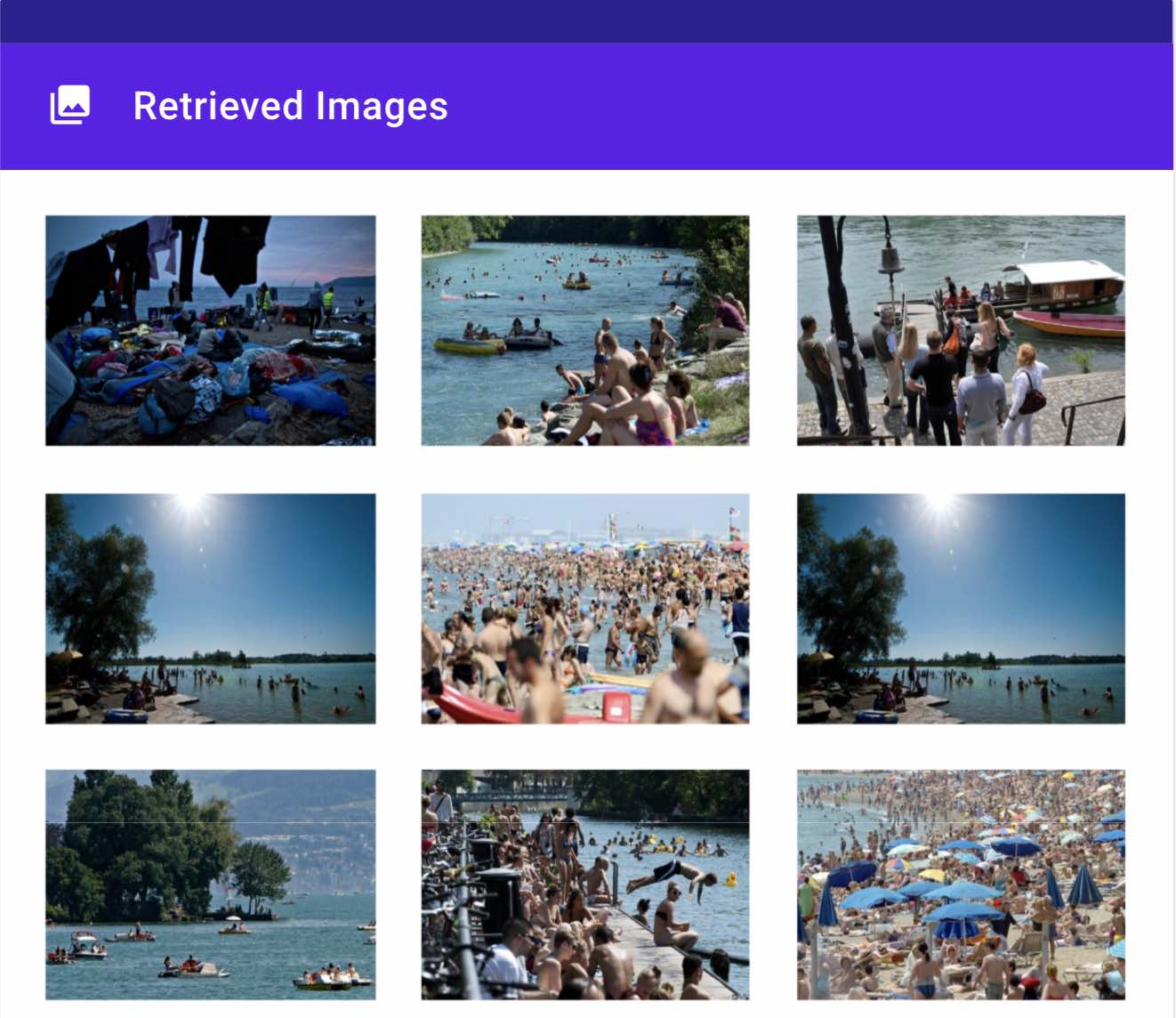}
\includegraphics[height=6.0cm]{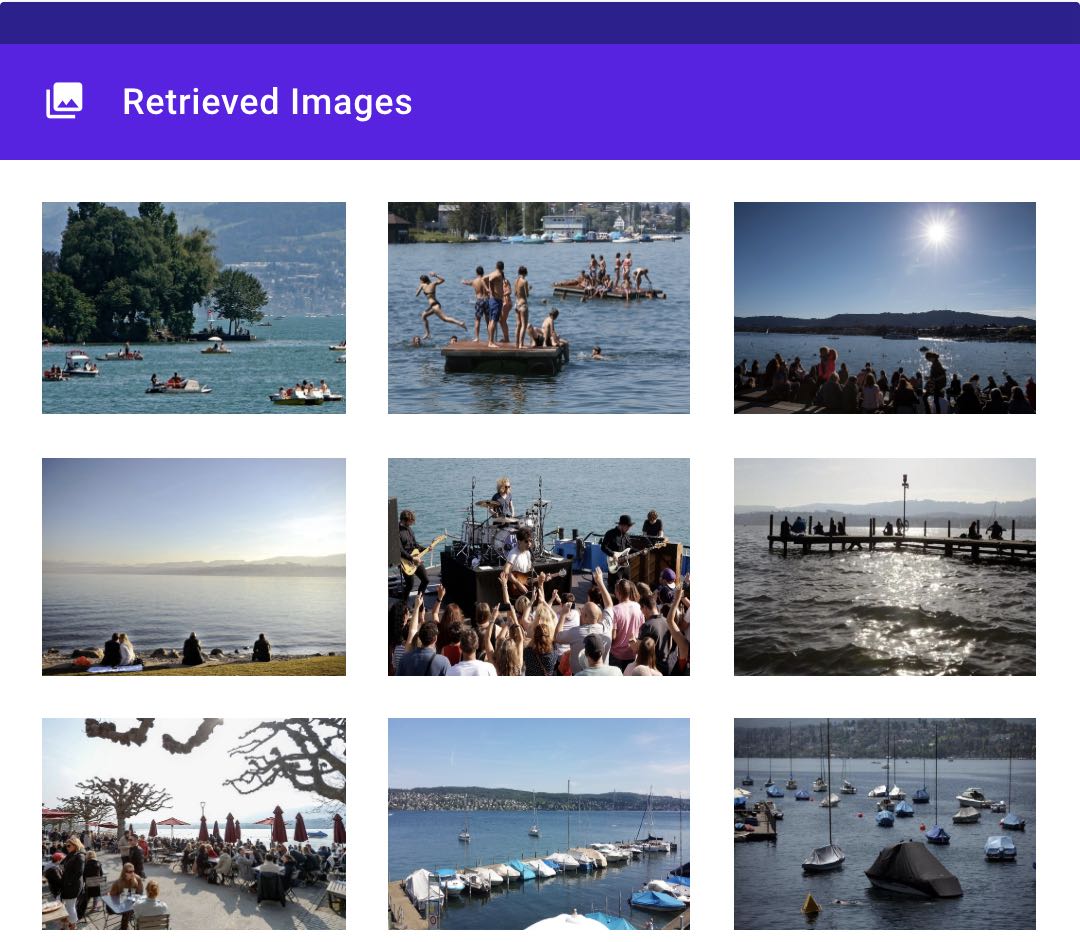}
\caption{Top 9 retrievals of the query ``\emph{Mehr Besucher am Z\"urichsee im Sommer.}'' as title. The left is without selecting any named entity; the right is searching with named entity ``Z\"urichsee''.}
\label{fig:lakezurich}
\end{figure}

Figure~\ref{fig:lakezurich} shows the top 9 retrievals with the German query ``\emph{Mehr Besucher am Z\"urichsee im Sommer.}''\footnote{English translation: \emph{More visitors coming to Lake Zurich in summer.}} as a caption. The left figure is searching without selecting any named entities. The retrievals appear to be people gathering around any water, not necessaily a lake or Lake Zurich. The right figure is searching with named entity ``Z\"urichsee''\footnote{English translation: Lake Zurich.}. Retrievals are still crowds and boats on/around waters but now we know that the water is guaranteed to be Lake Zurich.

\subsection{Naive Named Entity Filtering}

\begin{figure}[h]
\centering
\includegraphics[height=6.0cm]{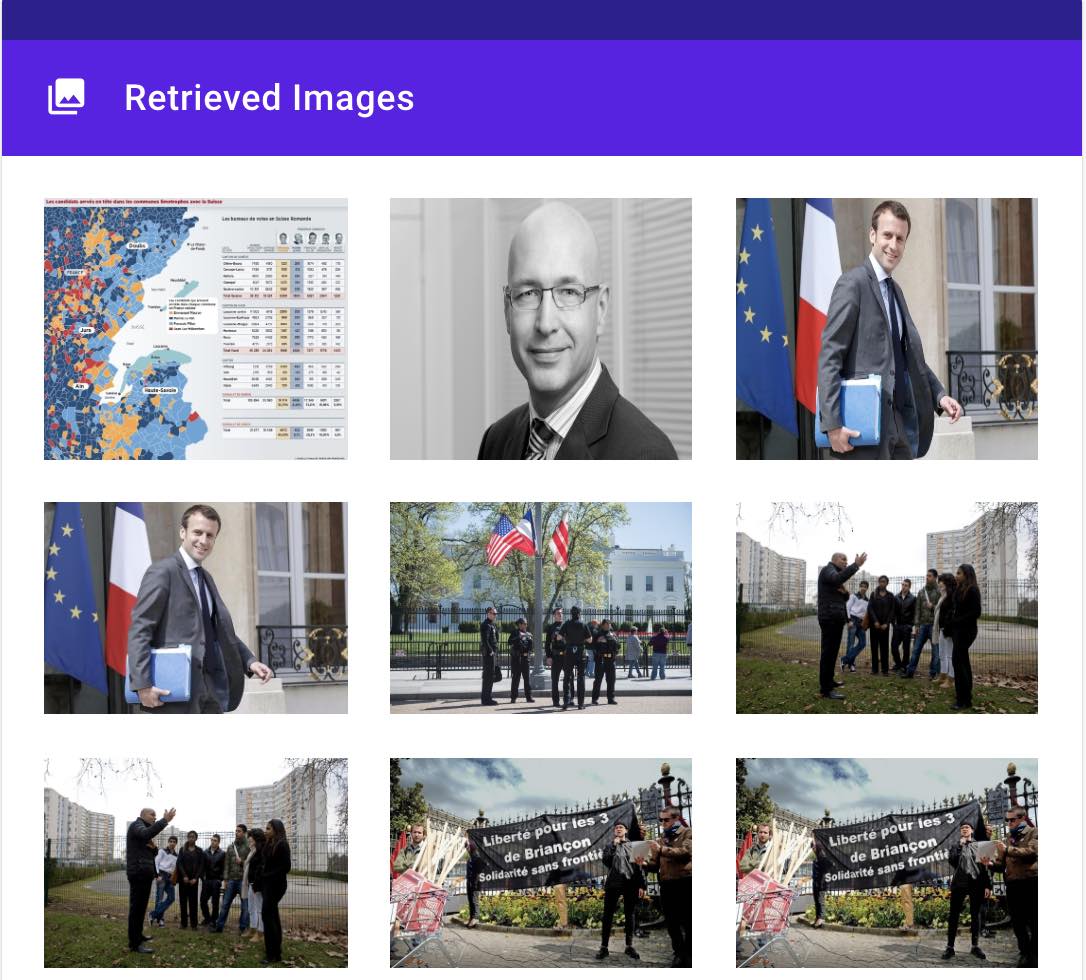}
\includegraphics[height=6.0cm]{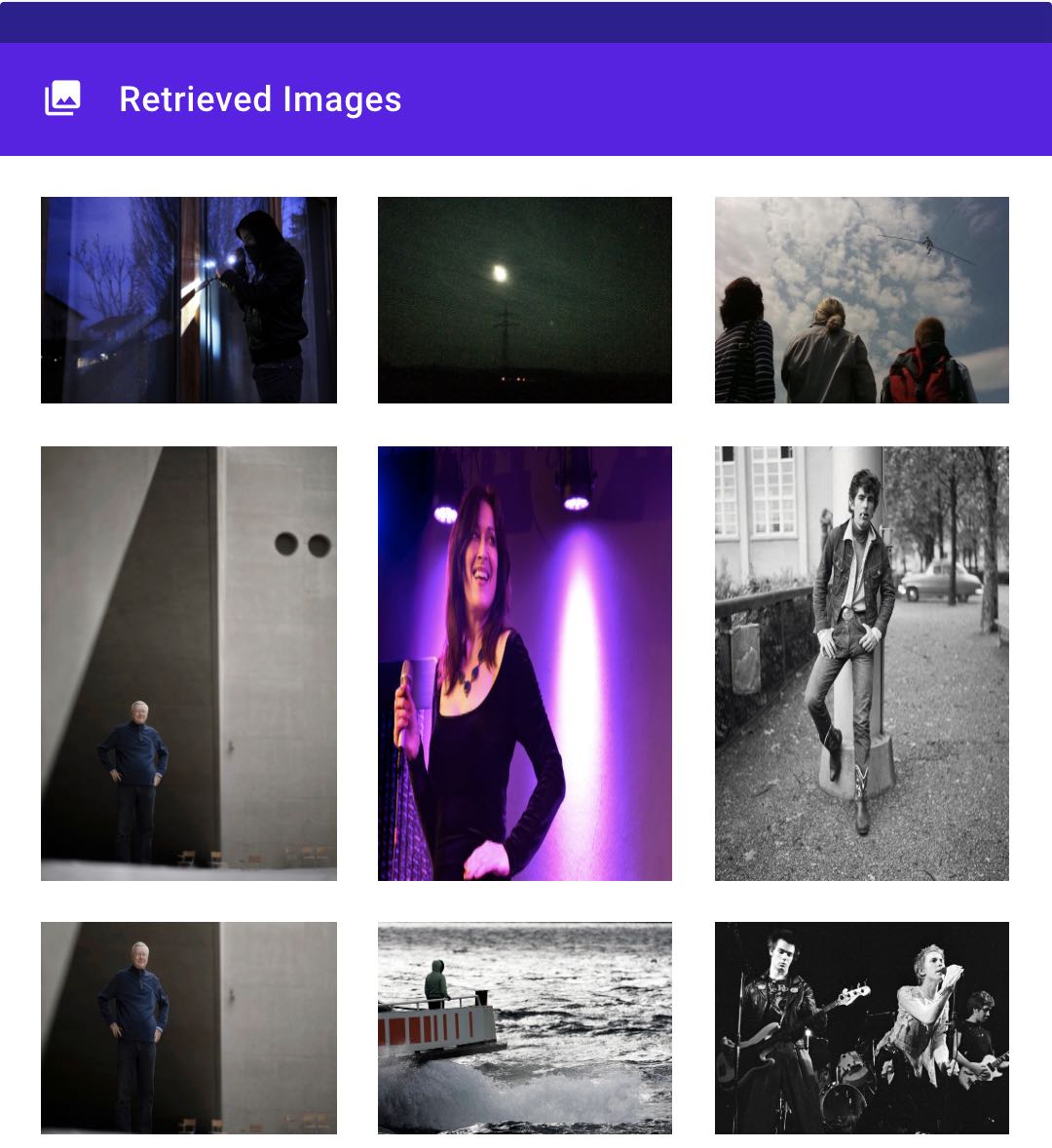}
\caption{Top 9 retrievals of searching with named entities only. Left is the result of ``macron'' and right is the results of ``Z\"urichsee''.}
\label{fig:naive_entity}
\end{figure}

One may wonder what happens when directly searching with named entities on metadata (without our model ranking the candidates). Here are the results in Figure~\ref{fig:naive_entity}. Only 2 of the 9 retrievals in the left figure has President Macron and the right figure looks completely random. It again verifies our claim that the metadata can be very heterogeneous and some might even be misleading.

